\documentclass[a4paper,showpacs,nofootinbib]{revtex4-1}
\usepackage[utf8]{inputenc}
\usepackage{graphicx,bm,color,amssymb,amsmath}
\usepackage{slashed,verbatim}
\usepackage{subfigure,array}
\usepackage[colorlinks=true,linktocpage=true,linkcolor=blue,citecolor=blue]{hyperref}

\def\nn {\nonumber}

\date{06/04/2019}

\begin{document}

\hspace{15cm}TIFR/TH/18-53\\

\title{Spectral function and dilepton rate from a strongly magnetised hot and dense medium in the light of mean field models}
\author{Chowdhury Aminul Islam}
\email{ca.islam@theory.tifr.res.in}
\affiliation{Department of Theoretical Physics, Tata Institute of Fundamental Research, Mumbai-400005, India.}
\author{Aritra Bandyopadhyay}
\email{aritrabanerjee.444@gmail.com}
\affiliation{Universidade Federal de Santa Maria (UFSM) Rio Grande de Sul, Brazil.}
\author{Pradip K. Roy}
\email{pradipk.roy@saha.ac.in }
\affiliation{Saha Institute of Nuclear Physics, 1/AF Bidhan Nagar, Kolkata-700064, India.}
\author{Sourav Sarkar}
\email{sourav@vecc.gov.in}
\affiliation{Variable Energy Cyclotron Centre, 1/AF Bidhan Nagar, Kolkata-700064, India.}

\begin{abstract}
We have calculated the electromagnetic spectral function (SF) vis-\`{a}-vis the dilepton rate (DR) by evaluating the one-loop photon polarisation tensor for a strongly magnetised hot and dense medium. The calculation is performed in the ambit of mean field models namely Nambu\textendash Jona-Lasinio (NJL) and its Polyakov loop extended version (PNJL) in the lowest Landau level approximation. These models allow for a dynamical generation of quark mass which, evidently, gets affected in the presence of a magnetised medium. We have shown that the strength of the SF gets boosted because of the presence of dynamical quarks in lieu of the current quarks. It increases as we increase the magnetic field for a given value of temperature in both NJL and PNJL models. This increment is further reflected in the DR which is enhanced as compared to the Born rate for certain values of invariant masses. We also discuss the impact of chemical potential on DR within the present scenario.
\end{abstract}

\maketitle

\section{Introduction}  

Understanding Quantum Chromodynamics (QCD) in presence of magnetic background has gained lots of contemporary research interests \cite{Kharzeev:2013jha}. It is important to study QCD in presence of external magnetic field not only  for  its  relevance with the astrophysical phenomena \cite{Lattimer:2006xb,Ferrer:2005vd,Ferrer:2006vw,Ferrer:2007iw,Fukushima:2007fc,Noronha:2007wg,Feng:2009vt} but also due to  the possibility of strong magnetic field generation in non-central heavy-ion collision (HIC) \cite{Skokov:2009qp} which sets the stage for investigation of magnetic modifications. Although the background fields  produced in Relativistic Heavy Ion Collider (RHIC) and Large Hadron Collider (LHC) are much smaller in comparison with  the  field strengths that prevailed during the cosmological electro-weak phase transition which may reach up to $eB\approx200m_\pi^2$ \cite{Vachaspati:1991nm}, they are strong enough ($eB\approx m_\pi^2$ at RHIC and $eB\approx 10m_\pi^2$ at LHC) to cast significant influence on the deconfined medium of strongly interacting quarks and gluons known as quark gluon plasma (QGP) which is created in such HICs\cite{Muller:1983ed,Heinz:2000bk}. So far a  large number of efforts has been put to understand the effects of this background magnetic field on the strongly interacting QGP. This results in a plethora of novel and interesting phenomena \textendash\, finite 
temperature magnetic catalysis~\cite{Alexandre:2000yf,Gusynin:1997kj,Lee:1997zj} and inverse magnetic 
catalysis~\cite{Bali:2011qj,Bornyakov:2013eya,Mueller:2015fka,Ayala:2014iba,Ayala:2014gwa,Ayala:2016sln,Ayala:2015bgv}; chiral magnetic effect~\cite{Kharzeev:2007jp,Fukushima:2008xe,Kharzeev:2009fn}, chiral-and color-symmetry broken/restoration phase~\cite{Fayazbakhsh:2010bh,Fayazbakhsh:2010gc}; modification of dispersion properties in a magnetised hot QED medium~\cite{Sadooghi:2015hha,Das:2017vfh,Karmakar:2018aig}, thermodynamic properties~\cite{Strickland:2012vu,Andersen:2014xxa,Bandyopadhyay:2017cle}; dilepton production from a hot magnetized QCD plasma~\cite{Tuchin:2012mf,Tuchin:2013bda,Tuchin:2013ie,Sadooghi:2016jyf,Bandyopadhyay:2016fyd,Bandyopadhyay:2017raf,Ghosh:2018xhh} to name a few.

We know that dynamical properties of many particle system can be investigated through the correlation function (CF)~\cite{Forster(book):1975HFBSCF,Callen:1951vq,Kubo:1957mj}. Thus, by calculating the correlation function and its spectral representation many of the hadronic properties can be studied. These kind of properties for QCD in vacuum have been well studied~\cite{Davidson:1995fq}. Now in presence of hot and dense medium such vacuum properties are modified~\cite{Kapusta_Gale(book):1996FTFTPA,Lebellac(book):1996TFT}, because the dispersion relation of the particles moving in the medium gets modified. In presence of background magnetic field there will be further modifications which necessitates the modification of the available theoretical tools. From the temporal and spatial parts of the CFs, separately, we can extract important informations regarding the medium. As for example, the temporal CF tells us about the response of the conserved density fluctuations, whereas the spatial CF provides information regarding the mass and width of the particle in a medium. Similarly, the spectral representation of the vector current-current CF is related to the production of lepton pair, which leaves the created medium with minimum interaction. Thus, the dilepton rate (DR) is considered as one of the most reliable probes for QGP.

Being an important probe, the rate of dilepton production in presence of magnetic field has been studied using different techniques. Some of the earliest studies are done by Tuchin~\cite{Tuchin:2012mf,Tuchin:2013bda,Tuchin:2013ie} where he used phenomenological input to estimate the DR. To obtain the DR~\cite{Tuchin:2013bda,Tuchin:2013ie} a semi-classical Weiszäcker-Williams method~\cite{Jackson(book):1975CE} is utilised. Employing formal field theoretical technique and using Ritus eigenfunction method DR from a magnetised hot and dense medium was calculated in~\cite{Sadooghi:2016jyf}. In two other articles, where two of the present authors were involved, such formal field theoretical techniques have been used along with Schwinger method~\cite{Schwinger:1951nm} to estimate the DR \textendash\, in one case with imaginary time formalism~\cite{Bandyopadhyay:2016fyd} and real time formalism in the other~\cite{Bandyopadhyay:2017raf}. The problem was revisited in the latter formalism in~\cite{Ghosh:2018xhh} for hot QCD matter and significant enhancement was found in the low mass region.

In all such studies the photon self energy\footnote{Dilepton production rate is related with the imaginary part of the photon self energy.\cite{Islam:2014sea}} is calculated applying a formal field theoretical treatment while using quark propagators modified by the background magnetic field. In this article we would like to estimate the DR in presence of a strong magnetic field within the ambit of effective models such as Nambu$\textendash$Jona-Lasinio (NJL) model~\cite{Nambu:1961tp,Nambu:1961fr,Klevansky:1992qe} 
and its extension with Polyakov loop (PNJL)~\cite{Fukushima:2003fw,Fukushima:2003fm,Ratti:2005jh}. Here we also extend our previous studies~\cite{Bandyopadhyay:2016fyd,Bandyopadhyay:2017raf} to incorporate a finite chemical potential in addition to temperature. Now, in NJL model the quark propagator will be modified by an effective mass which is dynamically generated in a magnetised hot and dense medium~\cite{Menezes:2008qt,Boomsma:2009yk,Chatterjee:2011ry,Avancini:2011zz,Farias:2014eca,Ferrer:2014qka,Yu:2014xoa,Mao:2016fha,Farias:2016gmy}. We shall see that this will have some impact on the DR that we wish to evaluate.

Ongoing investigations provide us a strong hint that the QGP created in the HICs is a strongly interacting one~\cite{Adler:2006yt,Adare:2006ti,Adare:2006nq,Abelev:2006db,Aamodt:2010pb,Aamodt:2010pa,Aamodt:2010jd,Aad:2010bu}. Such nonperturbative nature of the medium can be mimicked roughly, through PNJL model where a background gauge field is added to the usual NJL model. The properties of the magnetised medium can be affected by such nonperturbative behaviour. In one of our earlier study we have investigated this point and found that the spectral properties are influenced to quite an extent, specially the DR gets enhanced due to the nonperturbative effect through the Polyakov loop fields~\cite{Islam:2014sea}. This makes the present study more interesting where the effect of the strong magnetic field will be further influenced by the presence of a background gauge field, specifically through PNJL model augmented with magnetic field~\cite{Gatto:2010qs,Providencia:2014txa,Ferreira:2013tba,Ferreira:2014kpa}. We perform our calculation in the strong magnetic field limit, in which case only the lowest Landau level (LLL) remains significant. This could be relevant for the initial stages of HICs, when the magnetic field is supposed to be very strong.

The paper is organised as follows: In section~\ref{sec:model} we briefly introduce the (P)NJL models. The details of the calculation of SF and DR have been given in section~\ref{sec:cal}. Then in section~\ref{sec:res} we show the results obtained and finally conclude in section~\ref{sec:con}.

\section{Effective QCD Models}
\label{sec:model}
\subsection{(P)NJL Model}
Here we briefly describe the details of the effective models that we work with, namely the NJL and PNJL models. For the sake of completeness we start by recapitulating the case when the magnetic field is absent. The present work considers a two flavor NJL model.
The corresponding Lagrangian is
\cite{Nambu:1961tp,Nambu:1961fr,Klevansky:1992qe,Buballa:2003qv}
\begin{eqnarray}
\mathcal{L}_{\rm{NJL}} = \bar{\psi}(i\gamma_{\mu}\partial^{\mu}-m_0+\gamma_0\mu)\psi
+ \frac{G_{S}}{2}[(\bar{\psi}\psi)^{2}+(\bar{\psi}i\gamma_{5}\vec{\tau}\psi)^{2}],
\label{eqn:Lag_NJL}
\end{eqnarray}
where, $m_0 = $diag$(m_{u},m_{d})$ with $m_{u}=m_{d}$ and $\vec{\tau}$'s
are Pauli matrices. $ G_{S}$ is the coupling
constants for local scalar type four-quark interaction. 

 The scalar four quark interaction term leads to the formation
 of chiral condensate $\sigma=\langle \bar{\psi}\psi\rangle$.
 The thermodynamic potential in mean field approximation is given as:
 \begin{eqnarray}
\Omega_{\rm{NJL}}&=&\frac{G_{S}}{2}\sigma^2  +
\Omega_{\rm vac} \nonumber\\
 &-&2N_fN_cT\int\frac{d^3p}{(2\pi)^3} \left[{\rm{ln}}(1+e^{-(E_p-{\mu})/T)})+{\rm{ln}}(1+e^{-(E_p+{\mu})/T)})\right],
 \label{eqn:potential_NJL}
 \end{eqnarray}
where $\Omega_{\rm vac} = -
2N_fN_c\int_{\Lambda}\frac{d^3p}{(2\pi)^3}E_p$ is the vacuum energy with
$E_p=\sqrt{{\vec p}^2+M_f^2}$ being the energy of a quark with flavor $f$
having constituent mass or the dynamical mass $M_f$ and $\Lambda$ being a
finite three momentum cut-off. The values of $\Lambda\,(651\,{\rm MeV})$, $G_S\,(10.08\,{\rm GeV^{-2}})$ and $m_0\,(5.5\,{\rm MeV})$ 
are fitted to reproduce some of the QCD vacuum results~\cite{Ratti:2005jh}. The thermodynamic potential depends on the dynamical fermion
mass $M_f$ which depends on the scalar ($\sigma$) condensates through the relation
\begin{equation}
M_f=m_0-G_S\sigma.
\label{eqn:massgap}
\end{equation}

Let us now briefly discuss PNJL model \cite{Fukushima:2003fw,Fukushima:2003fm,Ratti:2005jh,Mukherjee:2006hq,Ghosh:2007wy}, 
where, in addition to those in the NJL model
we have a couple of additional mean fields in the form of the expectation
value of the Polyakov Loop fields   
$\Phi$ and its conjugate
$\bar{\Phi}$.
The Lagrangian for the two-flavor PNJL model is given by,
\begin{eqnarray}
 {\mathcal L}_{\rm PNJL} = \bar{\psi}(i\slashed D-m_0+\gamma_0\mu)\psi +
\frac{G_S}{2}[(\bar{\psi}\psi)^2+(\bar{\psi}i\gamma_5\vec{\tau}\psi)^2]-{\mathcal U}(\Phi[A],\bar{\Phi}[A],T),
\end{eqnarray}
where $D^\mu=\partial^\mu-ig{\mathcal A}^\mu_a\lambda_a/2$, 
${\mathcal A}^\mu_a$ being the $SU(3)$ background fields, $\lambda_a$'s
are the Gell-Mann matrices and ${\mathcal U}(\Phi,\bar{\Phi},T)$ is the effective Polyakov Loop gauge potential. The thermodynamic potential can be
obtained as
\begin{eqnarray}
{\Omega}_{\textrm{PNJL}} &=& \frac{ G_S}{2} \sigma^2 +\Omega_{\rm vac} \nonumber\\
 &-&2N_fT\int \frac{d^3p}{(2\pi)^3} \ln \left[1+ 3\left(\Phi +{\bar \Phi}
 e^{-(E_p-\mu)/T} \right)e^{-(E_p-\mu)/T} + e^{-3(E_p-\mu)/T} \right ]  \nonumber \\
 &-&  2N_fT\int \frac{d^3p}{(2\pi)^3} \ln \left[1+ 3\left({\bar \Phi} + \Phi
 e^{-(E_p+\mu)/T} \right)e^{-(E_p+\mu)/T} + e^{-3(E_p+\mu)/T} \right ] \nonumber \\ 
 &+&{\mathcal U}(\Phi,{\bar \Phi},T) -\kappa T^4 \ln[J(\Phi,{\bar \Phi})].
 \label{eqn:potential_PNJL}
\end{eqnarray}

The effective Polyakov Loop gauge potential is parametrized~\cite{Ratti:2005jh} as
\begin{equation}
 \frac{{\mathcal U}(\Phi,\bar{\Phi},T)}{T^4} = 
    -\frac{b_2(T)}{2}\Phi\bar{\Phi} -
    \frac{b_3}{6}(\Phi^3+{\bar{\Phi}}^3) +
    \frac{b_4}{4}(\bar{\Phi}\Phi)^2,
\label{eq.potential}
\end{equation}
with
\begin{equation}
 b_2(T) = a_0 + a_1\left(\frac{T_0}{T}\right) + a_2\left(\frac{T_0}{T}\right)^2 +
    a_3\left(\frac{T_0}{T}\right)^3. \nonumber\\
\end{equation}
  
Values of different coefficients $a_0,\ a_1,\ a_2,\ a_3,\ b_3$ , $b_4$ and $\kappa$ are same as those given in~\cite{Ghosh:2007wy, Hansen:2006ee}. $T_0$ is taken as 210 MeV. The Vandermonde (VdM) determinant $J(\Phi,{\bar \Phi})$ is given
as \cite{Ghosh:2007wy,Islam:2014tea}
\begin{equation}
J[\Phi, {\bar \Phi}]=\frac{27}{24\pi^2}\left[1-6\Phi {\bar \Phi}+
4(\Phi^3+{\bar \Phi}^3)-3{(\Phi {\bar \Phi})}^2\right].
\end{equation}

\subsection{In presence of magnetic field}
The Lagrangian of NJL model in presence of magnetic field reads as~\cite{Menezes:2008qt,Boomsma:2009yk,Chatterjee:2011ry,Avancini:2011zz,Farias:2014eca,Ferrer:2014qka,Yu:2014xoa,Mao:2016fha,Farias:2016gmy},
\begin{align}
\mathcal{L}_{\rm{NJL}}^B = \bar{\psi}(i\slashed D-m_0)\psi
+ \frac{G_{S}}{2}[(\bar{\psi}\psi)^{2}+(\bar{\psi}i\gamma_{5}\vec{\tau}\psi)^{2}]-\frac{1}{4}F^{\mu\nu}F_{\mu\nu},
\label{eqn:Lag_NJL_mag}
\end{align}
where $\slashed D=\gamma_\mu D^\mu$ and $D^\mu=\partial^\mu-iqA^\mu$ with $q$ being the electric charge ($q_u=2/3e$ and $q_d=-1/3e$; $e$ is the charge of a proton) and $A^\mu$ being the electromagnetic gauge field. The field strength tensor is given as $F^{\mu\nu}=\partial^\mu A^\nu-\partial^\nu A^\mu$. 
 
Now in presence of magnetic field the dispersion relation of quarks will be modified to
\begin{align}
E_f(B)=[M_f^2+p_z^2+(2l+1-s)\lvert q_f\rvert B]^{1/2},
\label{eqn:disp_mag}
\end{align}
where the magnetic field is taken to be in the $z$-direction. $l$ is the Landau level (LL) and $s$ is the spin states of the quark which we need to take care in presence of magnetic field. Note that now the particle energy depends on the flavour not only through $M_f$ but also through $q_f$. The magnetic field will also modify the integral over the three momenta as,
\begin{align}
\int\frac{d^3p}{(2\pi)^3}\rightarrow\frac{|q_f|B}{2\pi}\sum_{l=0}^{\infty}\int_{-\infty}^{\infty} \frac{dp_z}{2\pi}
\end{align}

Thus the thermodynamic potential in presence of magnetic field~\cite{Menezes:2008qt,Boomsma:2009yk,Chatterjee:2011ry,Avancini:2011zz,Farias:2014eca,Yu:2014xoa,Mao:2016fha,Farias:2016gmy} becomes
\begin{align}
\Omega_{\rm{NJL}}^B=&\frac{G_{S}}{2}\sigma^2-
\frac{N_c}{2\pi}\sum_{f,l,s}|q_f|B\int_{-\infty}^{\infty}\frac{dp_z}{(2\pi)}E_f(B) \nonumber\\
-&\frac{N_c}{2\pi}T\sum_{f,l,s}|q_f|B\int_{-\infty}^{\infty}\frac{dp_z}{(2\pi)} \left[{\rm{ln}}\left(1+e^{-(E_f(B)-{\mu})/T}\right)+{\rm{ln}}\left(1+e^{-(E_f(B)+{\mu})/T}\right) \right]+\frac{B^2}{2}.
\label{eqn:potential_NJL_mag}
\end{align}
Here $f$, $l$ and $s$ represent the sum over flavour, LL and the spin states, respectively. The magnetic field is applied in the $z$-direction and its magnitude is $B$, $\vec B=B \hat z$. The term $B^2/2$ in the thermodynamic potential arises from the background magnetic field. Note that the integral in the second term is ultraviolet divergent. This divergence will be taken care of through the three momentum cut-off ($\Lambda$) once we rewrite it in terms of $\Omega_{\rm vac}$ in equation~\ref{eqn:potential_NJL_mag_vac}. Also note that we can no more perform the flavour sum trivially as the expression depends on the magnitude of the quark charge. In fact one can separate out the up and down quark parts, since they couple to the magnetic field with different strength. However, although the quark condensates are different in presence of magnetic field we are still in the isospin-symmetric scenario, i.e., $m_u=m_d=m_0$. Thus the constituent quark masses for up ($M_u$) and down ($M_d$) quarks remain the same ($M_u=M_d=M$)~\cite{Boomsma:2009yk,Ferrari:2012yw}. It is given as $M=m_0-G_S\big(\sigma_u+\sigma_d\big)$. With this it is evident that the gap equations for up and down quarks will be the same. So we do not separate them here. Though one can treat them separately to see the isospin-symmetry breaking effects of unequal coupling strengths of up and down quarks to magnetic field~\cite{Boomsma:2009yk}. Also note that in presence of magnetic field the dispersion relation in equation~\ref{eqn:disp_mag} depends on the type of flavour only through the strength of the coupling to magnetic field and, expectedly, it will be different for different flavour except for the LLL.\\

In the same way we can write the PNJL model in presence of magnetic field~\cite{Gatto:2010qs,Providencia:2014txa,Ferreira:2013tba,Ferreira:2014kpa}, which reads
\begin{eqnarray}
 {\mathcal L}_{\rm PNJL}^B = \bar{\psi}(i\slashed D-m_0+\gamma_0\mu)\psi +
\frac{G_S}{2}[(\bar{\psi}\psi)^2+(\bar{\psi}i\gamma_5\vec{\tau}\psi)^2]-{\mathcal U}(\Phi[A],\bar{\Phi}[A],T)-\frac{1}{4}F^{\mu\nu}F_{\mu\nu},
\end{eqnarray}
where $\slashed D$ now also contains background gauge field along with the magnetic field with, $D^\mu=\partial^\mu-iq A^\mu-ig{\mathcal A}^\mu_a\lambda_a/2$. All the other notations carry their meanings as mentioned before. The corresponding thermodynamic potential~\cite{Gatto:2010qs,Providencia:2014txa,Ferreira:2013tba,Ferreira:2014kpa} is
\begin{eqnarray}
{\Omega}_{\textrm{PNJL}}^B &=& \frac{ G_S}{2} \sigma^2 -\frac{N_c}{2\pi}\sum_{f,l,s}|q_f|B\int_{-\infty}^{\infty}\frac{dp_z}{(2\pi)}E_f(B) \nonumber\\
 &-&\frac{1}{2\pi}T\sum_{f,l,s}|q_f|B\int_{-\infty}^{\infty}\frac{dp_z}{(2\pi)} \ln \left[1+ 3\left(\Phi +{\bar \Phi}
 e^{-(E_f(B)-\mu)/T} \right)e^{-(E_f(B)-\mu)/T} + e^{-3(E_f(B)-\mu)/T} \right ]  \nonumber \\
 &-&\frac{1}{2\pi}T\sum_{f,l,s}|q_f|B\int_{-\infty}^{\infty}\frac{dp_z}{(2\pi)}	 \ln \left[1+ 3\left({\bar \Phi} + \Phi
 e^{-(E_f(B)+\mu)/T} \right)e^{-(E_f(B)+\mu)/T} + e^{-3(E_f(B)+\mu)/T} \right ] \nonumber \\ 
 &+&{\mathcal U}(\Phi,{\bar \Phi},T) -\kappa T^4 \ln[J(\Phi,{\bar \Phi})]+\frac{B^2}{2}.
 \label{eqn:potential_PNJL_mag}
\end{eqnarray}
Here again the diverging second term will be rewritten in terms of $\Omega_{\rm vac}$ in equation~\ref{eqn:potential_PNJL_mag_vac} to render it finite. We can further simplify this potential using the technique used in~\cite{Menezes:2008qt} and write it in both the conditions \textendash~(i) $T=0,\,\mu\ne0\,\rm{and}\, B\ne0$ and (ii) all of them being nonzero. Here we concentrate only on the second case and the expression in equation~\ref{eqn:potential_NJL_mag} can further be written in terms of $\Omega_{\rm vac}$ as
\begin{align}
\Omega_{\rm{NJL}}^B=&\frac{G_{S}}{2}\sigma^2+
\Omega_{\rm vac}-\frac{N_c}{2\pi^2}\sum_{f}(|q_f|B)^2\big(\zeta^{\prime}(-1,x)+\frac{x^2}{4}-\frac{1}{2}(x^2-x){\rm ln}x\big) \nonumber\\
-&\frac{N_c}{2\pi}T\sum_{f,l,s}|q_f|B\int_{-\infty}^{\infty}\frac{dp_z}{(2\pi)} \left[{\rm{ln}}\left(1+e^{-(E_f(B)-{\mu})/T}\right)+{\rm{ln}}\left(1+e^{-(E_f(B)+{\mu})/T}\right) \right]+\frac{B^2}{2}.
\label{eqn:potential_NJL_mag_vac}
\end{align}
The corresponding PNJL potential is
\begin{align}
\Omega_{\rm{PNJL}}^B=&\frac{G_{S}}{2}\sigma^2+
\Omega_{\rm vac}-\frac{N_c}{2\pi^2}\sum_{f}(|q_f|B)^2\big(\zeta^{\prime}(-1,x)+\frac{x^2}{4}-\frac{1}{2}(x^2-x){\rm ln}x\big) \nonumber\\
-&\frac{1}{2\pi}T\sum_{f,l,s}|q_f|B\int_{-\infty}^{\infty}\frac{dp_z}{(2\pi)} \ln \left[1+ 3\left(\Phi +{\bar \Phi}
 e^{-(E_f(B)-\mu)/T} \right)e^{-(E_f(B)-\mu)/T} + e^{-3(E_f(B)-\mu)/T} \right ]  \nonumber \\
 -&\frac{1}{2\pi}T\sum_{f,l,s}|q_f|B\int_{-\infty}^{\infty}\frac{dp_z}{(2\pi)}	 \ln \left[1+ 3\left({\bar \Phi} + \Phi
 e^{-(E_f(B)+\mu)/T} \right)e^{-(E_f(B)+\mu)/T} + e^{-3(E_f(B)+\mu)/T} \right ] \nonumber \\ 
 +&{\mathcal U}(\Phi,{\bar \Phi},T) -\kappa T^4 \ln[J(\Phi,{\bar \Phi})]+\frac{B^2}{2}.
 \label{eqn:potential_PNJL_mag_vac}
\end{align}
Before we go to the calculation of vector CF in the effective models we discuss here a few of the intricacies. Some of the earliest studies using NJL model showed that a strong constant magnetic field enhances spontaneous chiral symmetry breaking through the generation of fermion dynamical mass~\cite{Gusynin:1995nb,Schramm:1991ex}. This phenomenon is known as magnetic catalysis (MC) and can be understood in terms of dimensional reduction of the system. This effect of MC was further obtained by many of the effective model studies~\cite{Boomsma:2009yk,Chatterjee:2011ry,Ferrari:2012yw,Ferreira:2013tba,Fraga:2008qn,Gatto:2010pt}, including NJL and PNJL. In models like PNJL behaviour of MC is found for both chiral and deconfinement dynamics. It is also found by different lattice QCD simulations that the values of light quark condensates increase at temperature well below and well above the critical temperature ($T_c$) but it decreases near $T_c$~\cite{Bali:2011qj,Bali:2012zg}. This behaviour of decreasing condensate values with the increase of magnetic field is dubbed as inverse magnetic catalysis (IMC). On the other hand IMC is counterintuitive and is probably occurring due to the dominance of sea contributions over the valence contributions of the condensate around $T_c$~\cite{Bruckmann:2013oba}. Once IMC is recognised, for both chiral and deconfinement transitions~\cite{Bruckmann:2013oba,Bornyakov:2013eya,Endrodi:2015oba} by LQCD calculation several attempts have been made to understand it through different effective QCD models~\cite{Mueller:2015fka,Farias:2014eca,Ferrer:2014qka,Yu:2014xoa,Mao:2016fha,Farias:2016gmy,Providencia:2014txa,Ferreira:2014kpa,Ayala:2014iba,Ayala:2014gwa,Ferrer:2014qka,Chao:2013qpa,Andersen:2014oaa}. For example the IMC phenomenon is observed in Ref~\cite{Farias:2014eca} through a magnetic field dependent coupling constant in NJL model, whereas within the same model similar behaviour is achieved in a strong magnetic field as the coupling constant becomes anisotropic and the quarks produce antiscreeing effect at LLL~\cite{Ferrer:2014qka}. At this point we also want to note that in the present work we will not be discussing about the effect of IMC because we will not be focussing in temperature ranges close to $T_c$, where IMC effects become dominant. Hence this study will only incorporate MC effects, as we will see later in our results.

\section{calculation}
\label{sec:cal}

\subsection{Vector correlation function in one-loop}
\label{sc.resum_corr}

\begin{figure} [!htb]
\center
\includegraphics[scale=0.4]{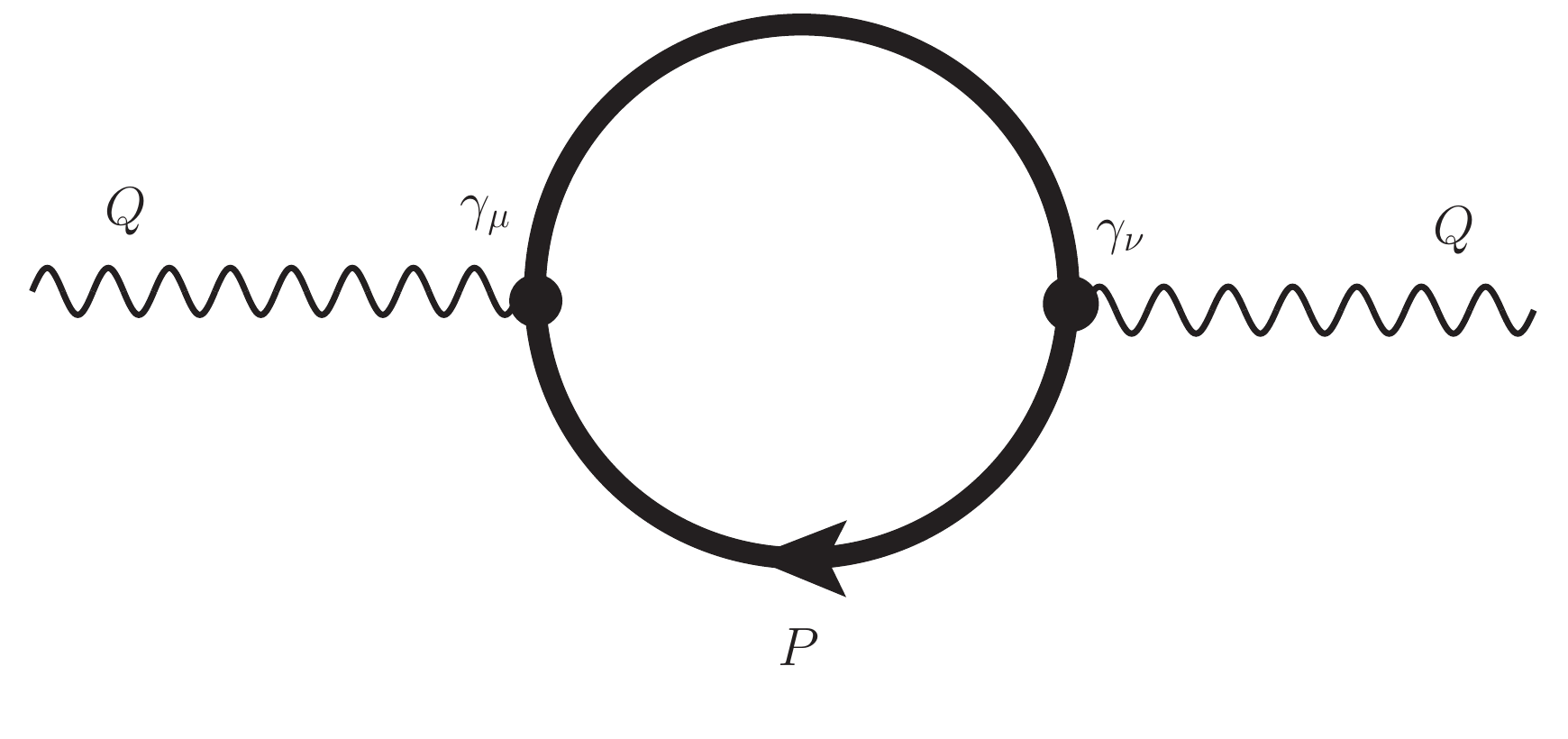}
\caption{Vector channel correlator at one-loop.}
\label{fg.corr_oneloop}
\end{figure}

In this section we would like to compute the current current correlator in NJL and PNJL models in presence of a strong external background magnetic field. The two point current-current correlator $C_{\mu\nu}(Q)$ is related to photon self-energy $\Pi_{\mu\nu}(Q)$ as 
\begin{eqnarray}
q_f^2C_{\mu\nu} (Q) &=&  \Pi_{\mu \nu} (Q), \label{corr_func}
\end{eqnarray}
with $q_f$ is the electric charge of a given quark flavour $f$. The photon self energy in one loop level (figure \ref{fg.corr_oneloop}) is again expressed as
\begin{eqnarray}
\Pi_{\mu\nu}(Q) = -i\sum_{f} q_f^2\int \frac{d^4k}{(2\pi)^4} 
\textsf{Tr}_{c}\left[\gamma_\mu S_m(P) 
\gamma_\nu S_m(K)\right] , \label{pola}
\end{eqnarray}
where $Q$ is the external momentum, $P$ and $K=P+Q$  are the loop 
momenta. $\textsf{Tr}_{c}$ represents both  color and Dirac traces whereas 
the $\sum_{f}$ is over flavor  because   
we have considered a two-flavor system ($N_f=2$)  of 
equal current quark mass ($m_f=m_u=m_d=5$ MeV if not said otherwise).
Finally the electromagnetic spectral representation is extracted from the imaginary part of the correlation function $C_\mu^\mu(Q)$ as
\begin{eqnarray}
\rho_V(Q) &=& \frac{1}{\pi} \mathcal{I}m~C^\mu_\mu(Q)=\frac{1}{\pi} 
\mathcal{I}m~\Pi^\mu_\mu(Q)/q_f^2.  \label{spec_func}
\end{eqnarray}

To write the fermionic propagator in presence of a background field we use the prescription provided by Schwinger~\cite{Schwinger:1951nm}. When the background magnetic field is very large ($q_fB\rightarrow\infty$), the only relevant LL is the lowest one. This point is well described through figure 1 as in~\cite{Bandyopadhyay:2016fyd}. In the strong field approximation the fermion propagator gets simplified to
\begin{align}
S(P)=e^{-{P_\perp^2}/{q_fB}}~~\frac{1}{\slashed{P}_\shortparallel-m_f}(1-i\gamma_1\gamma_2~{\rm sign}(q_fB)),
\label{eqn:prop_sfa}
\end{align}
where $P$ is the four momentum and $m_f$ and $q_f$ are the current mass and the absolute charge of the quarks, respectively. In presence of strong magnetic field there will be dynamical mass generation and the $m_f$ will be replaced by the effective quark mass. This is incorporated here through the effective model calculation. The notations that we use in equation~\ref{eqn:prop_sfa} and for the rest of the paper are as follows
\begin{align}
x^\mu =& x_\shortparallel^\mu + x_\perp^\mu;~~ x_\shortparallel^\mu = (x^0,0,0,x^3) ;~~  x_\perp^\mu = (0,x^1,x^2,0),\nn\\
g^{\mu\nu} =& g_\shortparallel^{\mu\nu} + g_\perp^{\mu\nu};~~ g_\shortparallel^{\mu\nu}= \textsf{diag}(1,0,0,-1);~~ g_\perp^{\mu\nu} = \textsf{diag}(0,-1,-1,0),\nn\\
(x\cdot y) =& (x\cdot y)_\shortparallel - (x\cdot y)_\perp;~~ (x\cdot y)_\shortparallel = x^0y^0-x^3y^3;~~ (x\cdot y)_\perp = (x^1y^1+x^2y^2),
\end{align}
where $\shortparallel$ and $\perp$ represent parallel and perpendicular components, respectively. Now the NJL quark propagator in Hartree approximation is written as
\begin{align}
\left[\slashed P -m_0+G_S\sigma\right ]^{-1}
= \left[\slashed P -M_f(B)\right ]^{-1},
\label{eqn:mod_prop_njl_inv}
\end{align}
where $M_f$ is the effective mass already defined in section~\ref{sec:model}; its magnetic field dependence, which arises through the gap equation, is now explicitly shown. We will elaborate the calculation for NJL model and the extension to PNJL one will be straightforward. Now we rewrite equation~\ref{eqn:prop_sfa} when the fermion mass is an effective one
\begin{align}
S_{\textrm{NJL}}(P)=e^{-{P_\perp^2}/{q_fB}}~~\frac{1}{\slashed{P}_\shortparallel-M_f(B)}(1-i\gamma_1\gamma_2~{\rm sign}(q_fB)).
\label{eqn:prop_sfa_eff}
\end{align}

So following equation~\ref{pola} the current-current correlator can be computed as
\begin{align}
\Pi_{\mu\nu}(Q)\Big\vert_{sfa} =& -i\sum_{f}q_f^2\int\frac{d^4P}{(2\pi)^4}\textsf{Tr}_c\left[\gamma_\mu S_{\textrm{NJL}}(P)\gamma_\nu S_{\textrm{NJL}}(K)\right]\nn\\
=& -iN_c\sum_{f}q_f^2 \int\frac{d^2P_\perp}{(2\pi)^2} \exp\left(\frac{-P_\perp^2-K_\perp^2}{q_fB}\right)\nn\\
&\times \int\frac{d^2P_\shortparallel}{(2\pi)^2} \textsf{Tr} \left[\gamma_\mu \frac{\slashed{P}_\shortparallel+M_f(B)}
{P_\shortparallel^2-M_f(B)^2}(1-i\gamma_1\gamma_2~{\rm sign}(q_fB))\gamma_\nu \frac{\slashed{K}_\shortparallel+M_f(B)}{K_\shortparallel^2-M_f(B)^2}(1-i\gamma_1\gamma_2~{\rm sign}(q_fB))\right],
\end{align}
where `\textit{sfa}' indicates the strong field approximation and $\textsf{Tr}$ represents the trace  in the Dirac space only. The longitudinal and transverse parts are now completely separated and we can perform the Gaussian integration over the transverse momenta, which leads to
\begin{align}
\Pi_{\mu\nu}(Q)\Big\vert_{sfa} = -iN_c\sum_{f}~e^{{-Q_\perp^2}/{2q_fB}}~~\frac{q_f^3 B}{\pi}\int\frac{d^2P_\shortparallel}{(2\pi)^2} 
\frac{S_{\mu\nu}}{(P_\shortparallel^2-M_f(B)^2)(K_\shortparallel^2-M_f(B)^2)}. 
\label{eqn:pol_vacuum}
\end{align}
The tensor structure $S_{\mu\nu}$ originates from the Dirac trace and is given by 
\begin{align}
S_{\mu\nu} = P_\mu^\shortparallel K_\nu^\shortparallel + K_\mu^\shortparallel P_\nu^\shortparallel 
- g_{\mu\nu}^\shortparallel \left((P\cdot K)_\shortparallel -M_f(B)^2\right),
\end{align}
where the Lorentz indices $\mu$ and $\nu$ can only take longitudinal values.

To evaluate the spectral function, we can contract the indices $\mu$ and $\nu$ in equation~\ref{eqn:pol_vacuum} which leads to a further simplification as
\begin{align}
\Pi_\mu^\mu(Q)\Big\vert_{sfa} =  -iN_c\sum_{f}~e^{{-Q_\perp^2}/{2q_fB}}~~\frac{q_f^3B}{\pi}\int\frac{d^2P_\shortparallel}{(2\pi)^2} 
\frac{2M_f(B)^2}{(P_\shortparallel^2-M_f(B)^2)(K_\shortparallel^2-M_f(B)^2)}. 
\end{align}

\subsubsection{Magnetised hot medium}
First we describe a strongly magnetised hot medium. To introduce temperature we use the imaginary time formalism technique and thus replace the $p_0$ integral by Matsubara sum as 
\begin{align}
\Pi_\mu^\mu(\omega,{\bf q})\Big\vert_{sfa} = -iN_c\sum_{f}~e^{{-Q_\perp^2}/{2q_fB}}~~\frac{2q_f^3BM_f(T,B)^2}{\pi}
\left(i T \sum_{p_0}\right)\int\frac{dp_3}{2\pi} 
\frac{1}{(P_\shortparallel^2-M_f(T,B)^2)(K_\shortparallel^2-M_f(T,B)^2)} \, . 
\label{eqn:pola_ft}
\end{align}
We should note here that the effective mass is now also dependent on temperature along with the magnetic field. One can always use the contour integration method to perform the Matsubara sum, but it becomes increasingly difficult when the number of propagators present in a diagrams increases. There is an elegant way to perform such complicated frequency sum using so called Saclay method. The essential trick, introduced by Pisarski~\cite{Pisarski:1987wc}, is to use propagator that are in co-ordinate representation in time but momentum representation in space: 
\begin{align}
\frac{1}{K_\shortparallel^2-M_f(T,B)^2} \equiv \frac{1}{k_0^2-E_k^2} = \int\limits_0^\beta d\tau e^{k_0\tau} \Delta_M(\tau,k),
\label{eqn:mixed_representation}
\end{align}
 and 
\begin{align}
\Delta_M(\tau,k) = \frac{1}{2E_k}\left[\left(1-n_F(E_k)\right)e^{-E_k\tau}-n_F(E_k)e^{E_k\tau}\right],
\end{align}
where $E_k=\sqrt{k_3^2+M_f(T,B)^2}$ and $n_F(x)=1/(\exp(\beta x)+1)$ is the Fermi-Dirac distribution function
with $\beta=1/T$. Using these, equation~\ref{eqn:pola_ft} can be simplified as
\begin{align}
\Pi_\mu^\mu(\omega,{\bf q})\Big\vert_{sfa} 
=& N_c\sum_{f}e^{\frac{-Q_\perp^2}{2q_fB}}~~\frac{2q_f^3BM_f(T,B)^2}{\pi}
T \sum_{k_0}\int\frac{dp_3}{2\pi} \int\limits_0^\beta~d\tau_1\int\limits_0^\beta~d\tau_2~ e^{p_0\tau_1}~e^{(p_0-q_0)\tau_2}\Delta_M(\tau_1,p)\Delta_M(\tau_2,k)\nn\\
=& N_c\sum_{f}e^{\frac{-Q_\perp^2}{2q_fB}}~~\frac{2q_f^3BM_f(T,B)^2}{\pi}\int\frac{dk_3}{2\pi}\int\limits_0^\beta~d\tau~ e^{q_0\tau}~\Delta_M(\tau,p)\Delta_M(\tau,k).
\end{align}
We now perform the $\tau$ integral to get
\begin{align}
\Pi_\mu^\mu(\omega,{\bf q})\Big\vert_{sfa}\!\! \!\! &=& \! \! 
N_c\sum_{f}e^{\frac{-Q_\perp^2}{2q_fB}}~~\frac{2q_f^3BM_f(T,B)^2}{\pi}\int\frac{dp_3}{2\pi}
\sum_{l,r=\pm 1}\!\!
\frac{\left(1-n_F(rE_p)\right)\left(1-n_F(lE_k)\right)}{4(rl)E_pE_k(p_0-rE_p-lE_k)}\left[e^{-\beta(rE_p+lE_k)}-1\right] ,
\label{eqn:Pi_sfa}
\end{align}
where the factors $r$ and $l$ can be related with the particles and antiparticles thereby representing various physical processes.
The discontinuity can be computed using
\begin{align}
\textsf{Disc~}\left[\frac{1}{\omega +\sum_i E_i}\right]_\omega = - \pi\delta(\omega + \sum_i E_i),
\label{disc_delta}
\end{align}
which leads to
\begin{align}
\mathcal{I}{ m}\, \Pi_\mu^\mu(\omega,{\bf q})\Big\vert_{sfa}\!\! \!\! =& \! \! 
-N_c \pi 
\sum_{f}e^{\frac{-Q_\perp^2}{2q_fB}}~~\frac{2q_f^3BM_f(T,B)^2}{\pi}\int\frac{dp_3}{2\pi}
\sum_{l,r=\pm 1}\!\!
\frac {\left(1-n_F(rE_p)\right)\left(1-n_F(lE_k)\right)}
{4(rl)E_pE_k} \nn \\
& \times \left[ e^{-\beta(rE_p+lE_k)}-1\right]
\delta(\omega-rE_p-lE_k).
\label{eqn:Pi_sfa_gen}
\end{align}

The general form of the delta function in equation~\ref{eqn:Pi_sfa_gen} corresponds to four processes~\cite{Bandyopadhyay:2016fyd}, depending on the values of $r$ and $l$. Among them, $r=1$ and $l=1$ corresponds to a process where a quark and a antiquark 
annihilate to a virtual photon, which is the only allowed process.
So, for this case, one can write from equation~\ref{eqn:Pi_sfa_gen},
\begin{align}
\mathcal{I}m~\Pi_\mu^\mu(\omega,{\bf q})\Big\vert_{sfa} =& 
N_c \pi \sum_{f}
e^{\frac{-Q_\perp^2}{2q_fB}}~~\frac{2q_f^3BM_f(T,B)^2}{\pi}
\int\frac{dp_3}{2\pi}~\delta(\omega-E_p-E_k)\frac{\left[1-n_F(E_p)-n_F(E_k)\right]}{
4E_pE_k}.
\label{eqn:Pi_im}
\end{align}
After performing the $p_3$ integral, the spectral function  in strong field approximation is obtained as
\begin{align}
\rho_V^1(T,B) \Big\vert_{sfa}=&\frac{1}{\pi}\mathcal{I}m~ 
C^\mu_\mu (Q)\Big\vert_{sfa}\nn\\
=& N_c\sum_{f}\frac{q_fBM_f(T,B)^2}{\pi^2 Q_\shortparallel^2}~e^{-{Q_\perp^2}/{2q_fB}}~\Theta
\left(Q_\shortparallel^2-4M_f(T,B)^2\right)\left(1-\frac{4M_f(T,B)^2}{Q_\shortparallel^2}
\right)^{-{1}/{2}} \Bigl[1-n_F(q_+^s)-n_F(q_-^s)\Bigr],
\label{eqn:spec_sfa}
\end{align}
where 
\begin{align}
q_\pm^s = \frac{\omega}{2}\pm \frac{q_3}{2}\sqrt{\left(1-\frac{4M_f(T,B)^2}{Q_\shortparallel^2}\right)},
\end{align}
with the superscript $s$ representing the strong magnetic field approximated result.

\subsubsection{Magnetised hot and dense medium}
In this section we incorporate the effect of baryon density as well in order to describe the general case of a strongly magnetised hot and dense medium. Here the effective mass further becomes dependent on chemical potential along with $T$ and $B$ through the gap equation. In presence of chemical potential the NJL propagator~(\ref{eqn:mod_prop_njl_inv}) gets modified to
\begin{align}
\left[\slashed P -M_f(T,\mu,B) +\gamma_0{\mu}\right ]^{-1}.
\label{eqn:mod_prop_njl_den_inv}
\end{align}
So the Matsubara sum includes chemical potential and gets modified to $\sum_{k_0=(2n+1)i\pi T}f(k_0)\rightarrow\sum_{k_0=(2n+1)i\pi T-\mu}f(k_0)$. This replacement eventually will have effects only in the distribution function. Thus equation~\ref{eqn:spec_sfa} gets modified to
\begin{align}
\rho_V^{2}(T, \mu, B)\Big\vert_{sfa}=&\frac{1}{\pi}\mathcal{I}m~ 
C^\mu_\mu (Q)\Big\vert_{sfa}\nn\\
=& N_c\sum_{f}\frac{q_fBM_f(T,\mu,B)^2}{\pi^2 Q_\shortparallel^2}~e^{-{Q_\perp^2}/{2q_fB}}~\Theta
\left(Q_\shortparallel^2-4M_f(T,\mu,B)^2\right)\left(1-\frac{4M_f(T,\mu,B)^2}{Q_\shortparallel^2}
\right)^{-{1}/{2}}\nonumber\\
&\Bigl[1-\frac{1}{2}\big(n_F^-(q_+^s)+n_F^+(q_+^s)\big)-\frac{1}{2}\big(n_F^-(q_-^s)+n_F^+(q_-^s)\big)\Bigr],
\label{eqn:spec_sfa_gen}
\end{align}
where $n_F^\mp(q^s)=n_F(q^s\mp\mu)$. We use equation~\ref{eqn:spec_sfa_gen} to calculate SF and also use it to get DR in the ambit of NJL model. It expectedly reduces to equation~\ref{eqn:spec_sfa} for $\mu=0$.

\subsubsection{Magnetised hot and dense medium in presence of Polyakov loop field}
We now want to extend our study by including a background gauge field (chromo). This is done through the PNJL model for which the effective propagator reads as
\begin{align}
\left[\slashed P -M_f(T,\mu,B, \Phi) +\gamma_0{\mu}-i\gamma_0{\cal A}_4\right]^{-1}, 
\label{eqn:mod_prop_pnjl_inv}
\end{align}
where ${\cal A}_4$ is the background gauge field. One crucial difference with the NJL model is that now the effective mass also varies with ${\cal A}_4$. This dependence is denoted through the Polyakov loop ($\Phi$). The whole procedure of calculation will remain the same, but now with $A_4$ the colour trace will be nontrivial. In usual NJL model the colour trace just gives a factor of $N_c$. Whereas in PNJL model the colour trace modifies the usual Fermi-Dirac distribution functions. The distribution functions of particle and anti-particle get modified to~\cite{Hansen:2006ee,Islam:2014sea}
\begin{subequations}
\begin{align}
f_-&=\frac{\Phi e^{-\beta(E_k- \mu)}
 +2\bar{\Phi}e^{-2\beta(E_k- \mu)}+e^{-3\beta(E_k- \mu)}}
 {1+3\Phi e^{-\beta(E_k-\mu)}+3\bar{\Phi}e^{-2\beta(E_k-\mu)}
 +e^{-3\beta(E_k-\mu)}}\\[1ex]
&{\rm and}\nonumber\\[1ex]
f_+&=\frac{\bar\Phi e^{-\beta(E_k+ \mu)}
 +2{\Phi}e^{-2\beta(E_k+ \mu)}+e^{-3\beta(E_k+ \mu)}}
 {1+3\bar\Phi e^{-\beta(E_k+\mu)}+3{\Phi}e^{-2\beta(E_k+\mu)}
 +e^{-3\beta(E_k+\mu)}},
\end{align}
\label{eqn:dis_modified}
\end{subequations}
respectively. $f_-$ and $f_+$ represent the distribution functions for particle and antiparticle respectively. 
 Thus equation~\ref{eqn:spec_sfa} gets modified to
\begin{align}
\rho_V^{3}(T, \mu, B, \Phi)\Big\vert_{sfa}
=& N_c\sum_{f}\frac{q_fBM_f(T,\mu,B, \Phi)^2}{\pi^2 Q_\shortparallel^2}~e^{-{Q_\perp^2}/{2q_fB}}~\Theta
\left(Q_\shortparallel^2-4M_f(T,\mu,B, \Phi)^2\right)\left(1-\frac{4M_f(T,\mu,B, \Phi)^2}{Q_\shortparallel^2}
\right)^{-{1}/{2}}\nonumber\\
&\Bigl[1-\frac{1}{2}\big(f_-(q_+^s)+f_+(q_+^s)\big)-\frac{1}{2}\big(f_-(q_-^s)+f_+(q_-^s)\big)\Bigr].
\label{eqn:spec_sfa_gen}
\end{align}

With $\Phi=\bar\Phi=1$ we get back the usual Fermi-Dirac distribution function. This essentially happens when we go to a temperature much higher than the critical temperature. For $\mu=0$, we have $\Phi=\bar\Phi$ which implies $f_-=f_+$. So, now we have spectral function for three different cases: (a) only magnetised hot medium denoted as $\rho_V^{1}$, (b) magnetised hot and dense medium is denoted as $\rho_V^{2}$ and (c) magnetised hot and dense medium in presence of Polyakov loop field is denoted as $\rho_V^{3}$.

\subsection{Dilepton rate}
Once we have the vector spectral function ($\rho_V$, in general) we can calculate the dilepton rate. Now in presence of magnetic field there are three possibilities: (a) only the initial quark-pair is moving in presence of the magnetic field, (b) both the initial quark-pair and the final lepton-pair are moving in the magnetised medium and (c) only the final lepton-pair gets influenced by the magnetic field. We will consider here both the cases (a) and (b). Case (a) is the most relevant scenario for a non-central heavy-ion collisions where the magnetic field is decaying very fast~\cite{Bzdak:2012fr,McLerran:2013hla}. 

For case (a) the dilepton production rate for massless ($m_l=0$) leptons with $N_f=2$ can be written from \cite{Bandyopadhyay:2016fyd} as
\begin{eqnarray}
\frac{dR}{d^4xd^4Q} &=& 
 \frac{5\alpha_{\mbox{em}}^2}{27\pi^2} 
\frac{n_B(q_0)}{Q^2} \rho_V(T,\mu,B,\Phi), 
\label{dcasea}
\end{eqnarray}

Similarly for case (b) the dilepton production rate with $N_f=2$ can be written from \cite{Bandyopadhyay:2016fyd} as
\begin{eqnarray}
\frac{dN^m}{d^4xd^4Q} &=& 
\frac{10\alpha_{\mbox{em}}^2}{9\pi^2}\frac{n_B(q_0)}{Q_\shortparallel^2Q^4}
~|eB| m_l^2\left(1-\frac 
{4m_l^2}{Q_\shortparallel^2}\right)^{-{1}/{2}}
\rho_V(T,\mu,B,\Phi).
\label{dcaseb}
\end{eqnarray}

Detailed discussions on different scenarios can be found in these articles~\cite{Tuchin:2013bda,Bandyopadhyay:2016fyd}.

\section{Results}
\label{sec:res}
\subsection{Effective models}
\begin{figure}[hbt]
\begin{center}
\subfigure[]
{\includegraphics[scale=0.35]{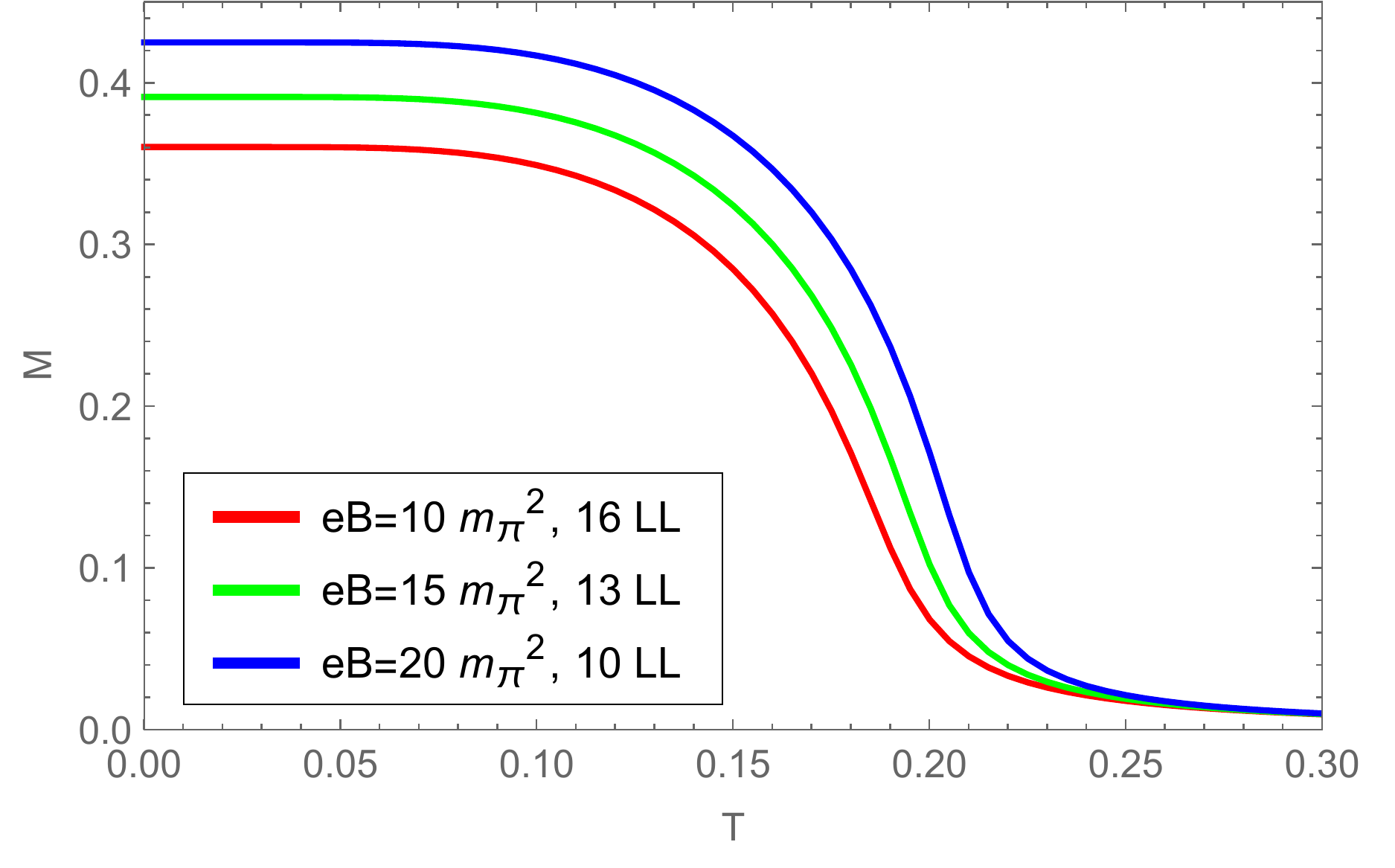}
\label{fig:M_nonzeroB}}
\subfigure[]
{\includegraphics[scale=0.35]{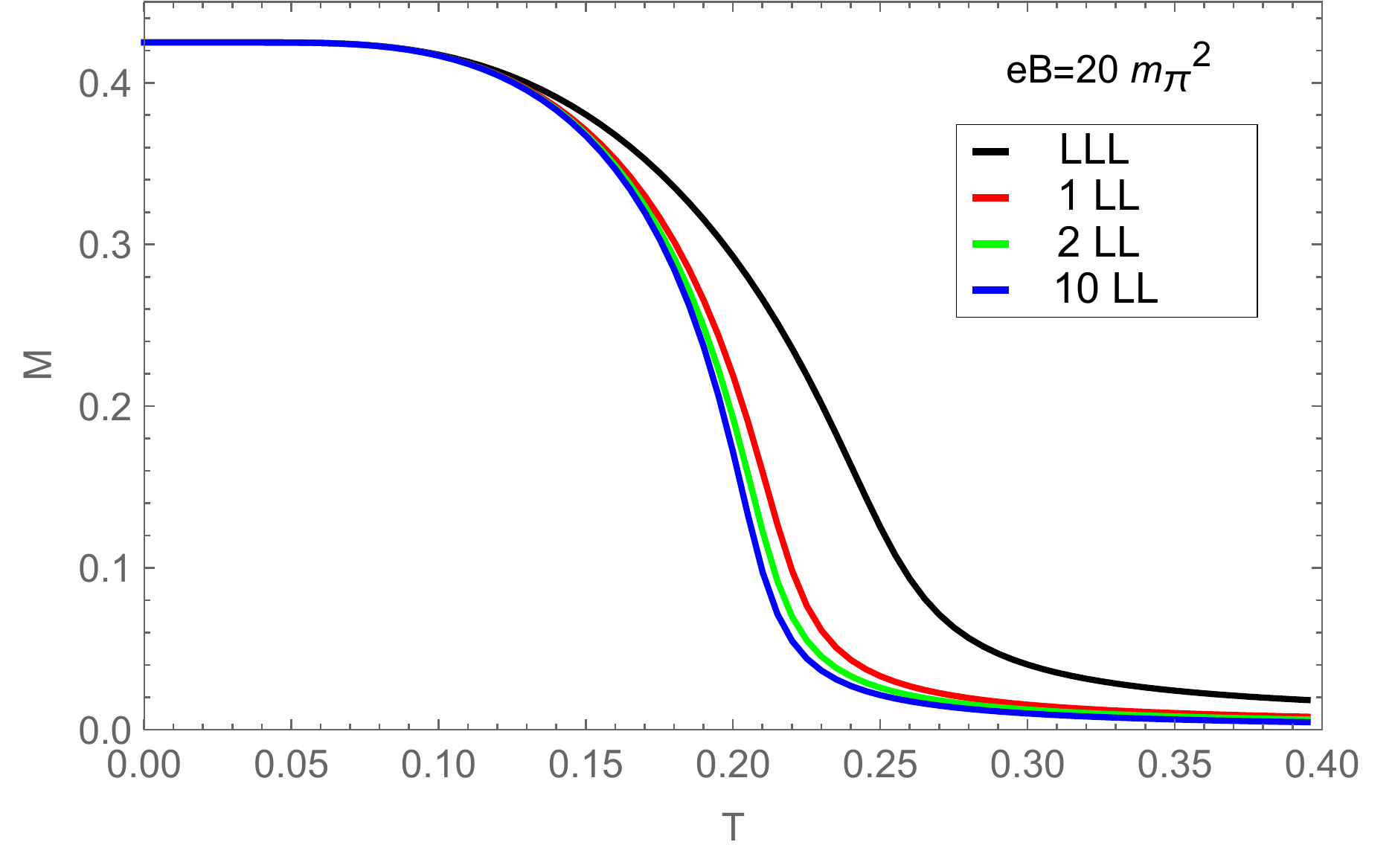}
\label{fig:MvsT_diffLL}}
\end{center}
\caption{NJL: In the left panel the plot of the constituent mass as a function of $T$ for different values of $B$ is shown. In the right panel it shows the difference in contribution while considering different number of Landau levels.}
\label{fig:M_njl}
\end{figure}

We have the plots for the constituent mass for different values of $B$ in the nonzero quark mass limit for NJL model in figure~\ref{fig:M_njl}. In the left panel~\ref{fig:M_nonzeroB} we have shown how the constituent mass increases with the increase of the magnetic field (i.e. Magnetic Catalysis (MC) effect). Depending on the value of the magnetic field generally one needs to sum up appropriate number of Landau levels. As in our main results of spectral properties, we will be using LLL approximation; in the right panel~\ref{fig:MvsT_diffLL} we have also shown the validity of the LLL approximation as a function of $T$ for a given magnetic field. The choice of LLL and its validity are discussed in numerical details at the appropriate places.

\begin{figure}[hbt]
\begin{center}
\subfigure[]
{\includegraphics[scale=0.35]{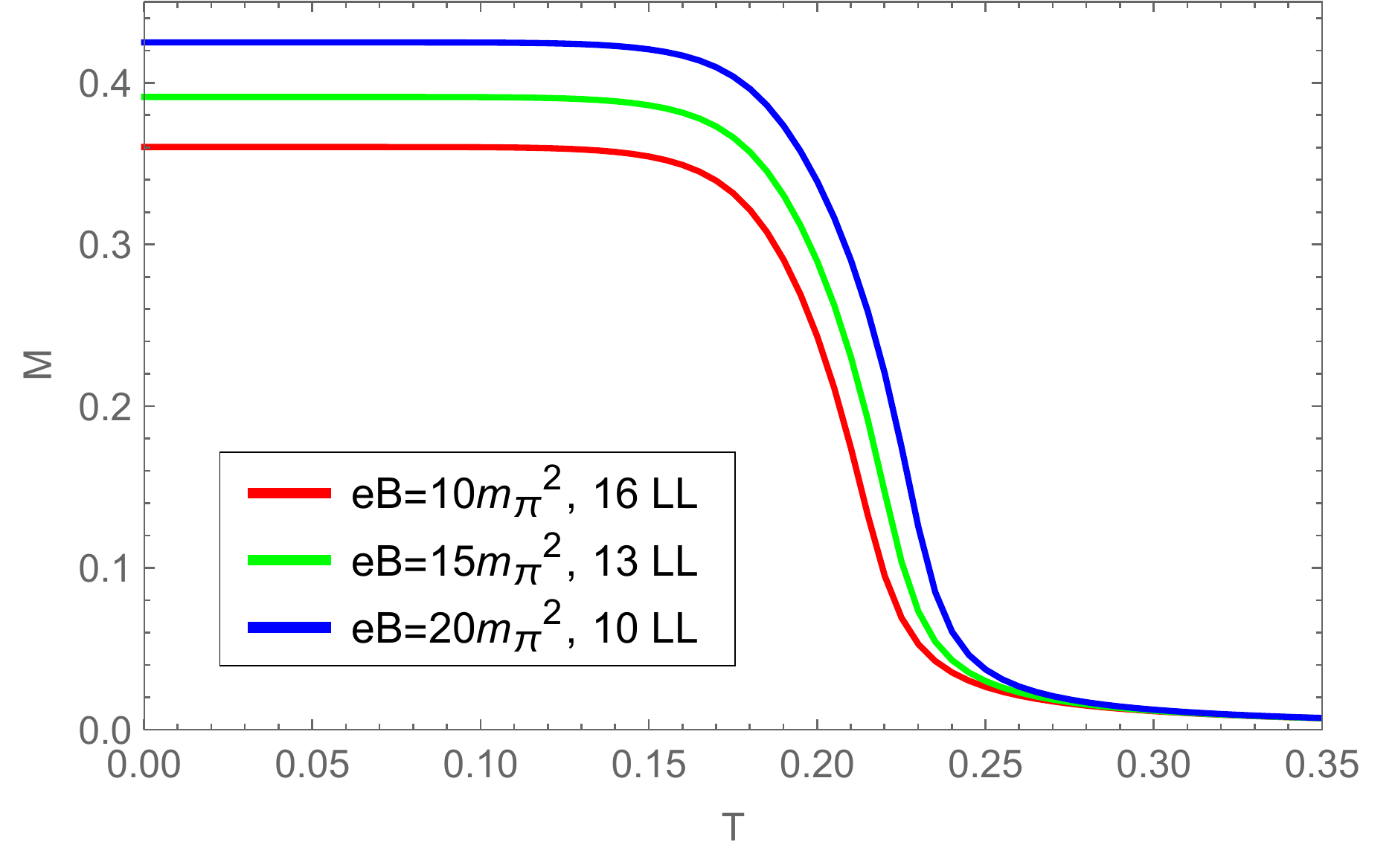}
\label{fig:M_nonzeroB_pnjl}}
\subfigure[]
{\includegraphics[scale=0.35]{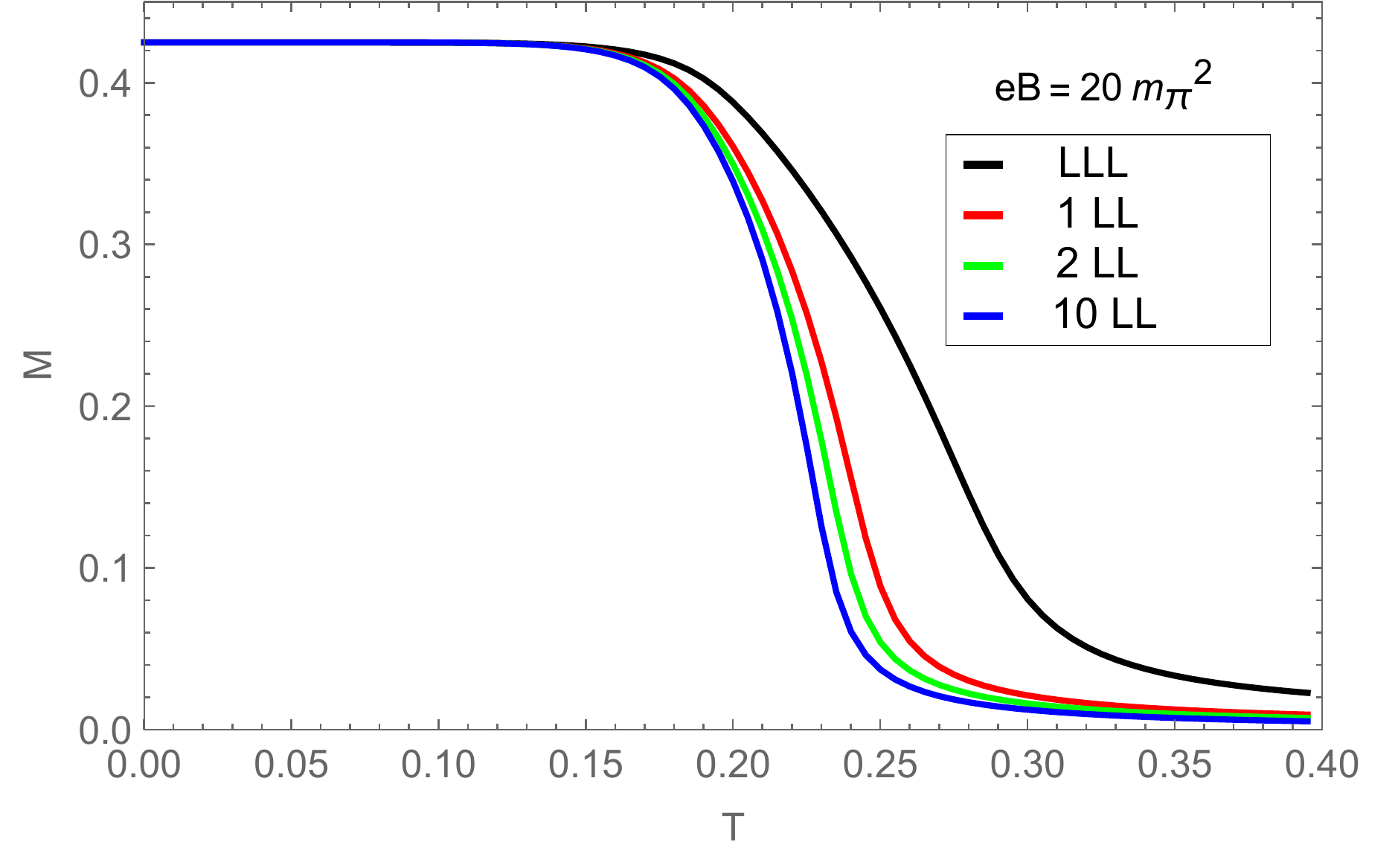}
\label{fig:M_pnjl_diffLL}}
\end{center}
\caption{PNJL: In the left panel the plot of the constituent mass as a function of $T$ for different values of $B$ is depicted. In the right panel it shows the difference in contribution while considering different number of Landau levels.}
\label{fig:M_pnjl}
\end{figure}

\begin{figure}[hbt]
\begin{center}
\subfigure[]
{\includegraphics[scale=0.35]{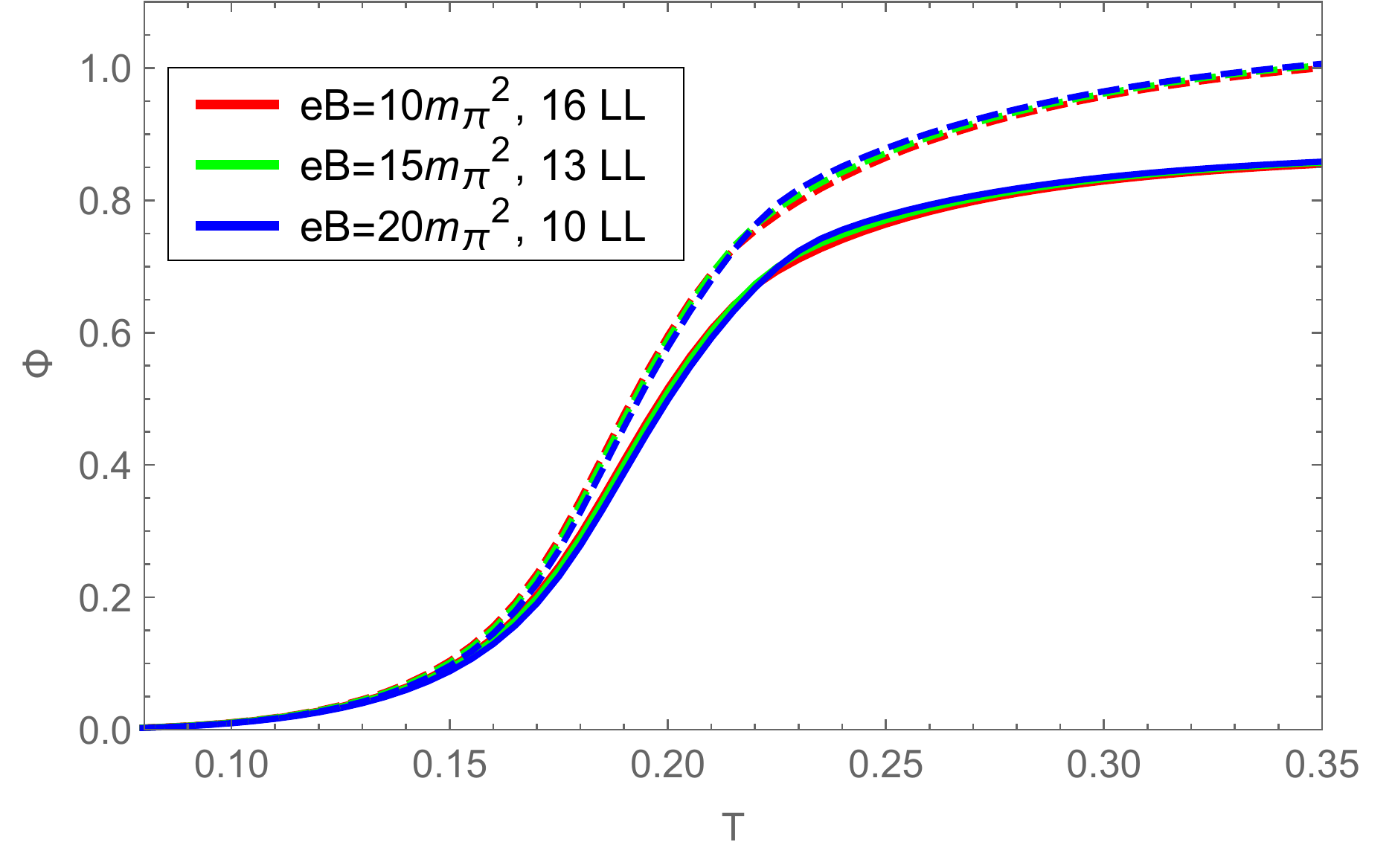}
\label{fig:phivsT_com}}
\subfigure[]
{\includegraphics[scale=0.35]{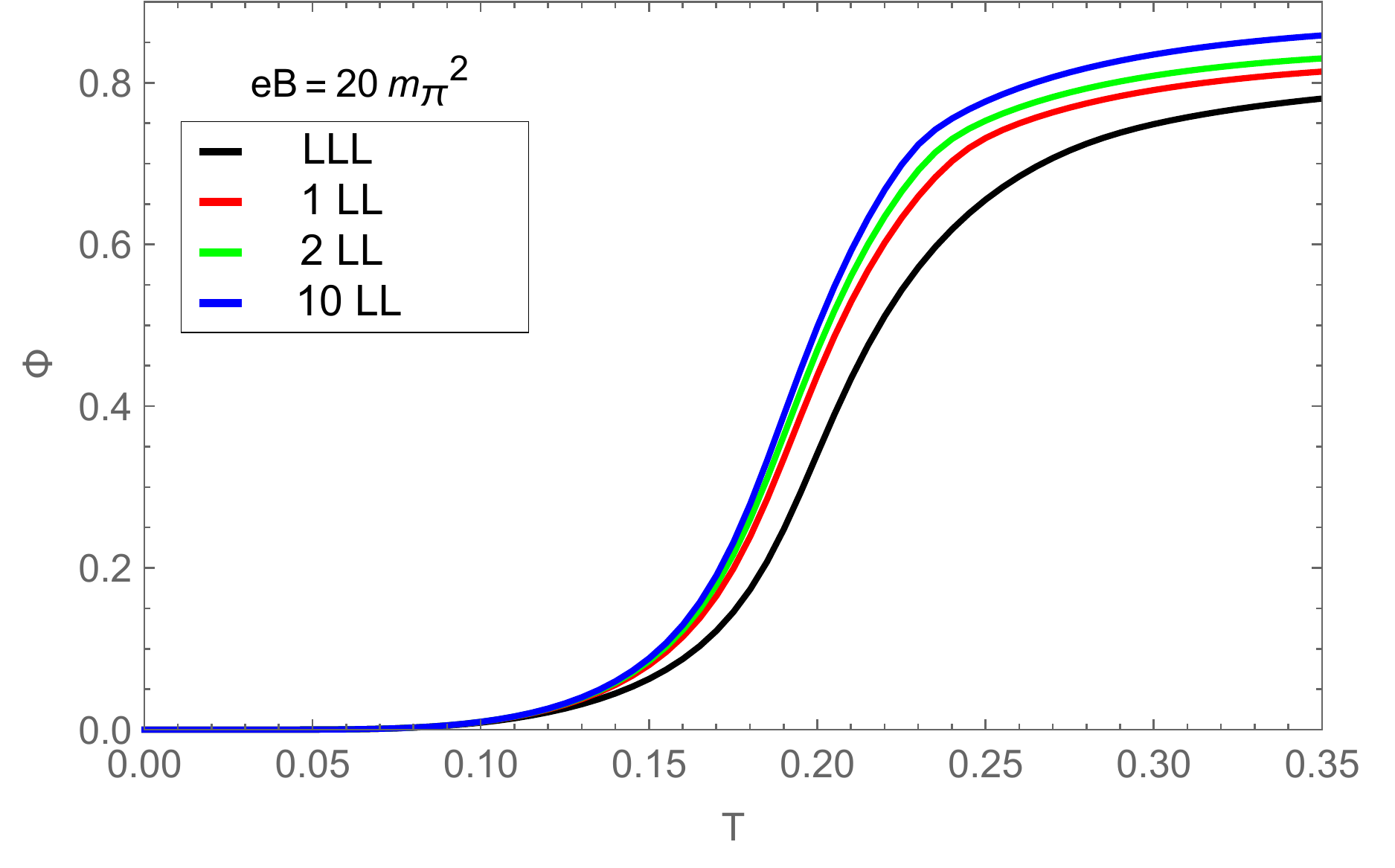}
\label{fig:phivsT_diffLL}}
\end{center}
	\caption{PNJL: Comparison between Polyakov loops with (solid line) and without (dashed line) the VdM term (left panel). Right panel shows the difference in contribution while considering different number of Landau levels.}
\label{fig:phivsT}
\end{figure}

Similar plots for the PNJL model are given in figures~\ref{fig:M_pnjl} and~\ref{fig:phivsT}. Along with the variation of the constituent mass shown in figure~\ref{fig:M_pnjl}, as an essential part of the PNJL model we have also plotted the Polyakov loop field as a function of temperature in figure~\ref{fig:phivsT}. In the left panel~\ref{fig:phivsT_com} the Polyakov loop has been plotted for different values of magnetic field. Here also we get the usual MC effect, though the catalysing effects are relatively mild. In the same plot we have also shown the difference in results between the presence and the absence of Vandermonde (VdM) term in the PNJL Lagrangian emphasising the importance of the inclusion of the same. If VdM term is not considered then as shown in figure~\ref{fig:phivsT_com} the Polyakov loop exceeds the value of unity at high temperature, which is not desired\footnote{We know that Polyakov loop $\Phi$ (its conjugate $\bar\Phi$) is the normalised trace of the Wilson line $\bf L$ (${\bf L}^\dagger$). Since the Wilson line itself is an SU(3) matrix, $\Phi(\bar\Phi)$ should obey the condition $0\le\Phi,\bar\Phi\le1$. Constraining the Polyakov loop within the limit $[0,1]$ does not have much impact on thermodynamic quantities like pressure, energy density, specific heat etc, but it helps for quantities like different susceptibilities to approach appropriate limit at high temperatures~\cite{Ghosh:2007wy}. We should also mention here that the VdM kind of term is not necessary for every form of the Polyakov loop potential, for example the form given in~\cite{Roessner:2006xn}, which always keeps the value of Polyakov loop within unity.}. We observed that the VdM term affects the transition temperatures for both chiral and deconfinement phase transitions. With the inclusion of it the chiral transition is increased by $5-6$ MeV and the deconfinement transition by $3-4$ MeV for different values of magnetic field that we have investigated (vide table~\ref{tab:tran_temp}). In all our later results we have kept the VdM term. In the right panel~\ref{fig:phivsT_diffLL} we have shown the validity of LLL approximation for the temperature variation of Polyakov loop field for a given strength of magnetic field.

\begin{table}[h]
\begin{center}
\begin{tabular}{c|ccccc|c}
\hline
$eB\,(m_\pi^2)$  & & $5$ & $10$ & $15$  &$20$& \\ \hline
$T_\sigma$ (MeV) & Without VdM & 204  & 208  & 214 & 221 \\ 
 & With VdM & 209  & 213  & 220 & 227 \\ \hline
$T_\Phi$ (MeV) & Without VdM & 184.9  & 185.2  & 185.7 & 186.3 \\ 
 & With VdM & 187.9  & 188.2  & 188.8 & 189.5 \\ \hline 
\end{tabular}
\end{center}
\caption{Temperatures for chiral and deconfinement phase transitions as a function of magnetic field for both in absence and presence of VdM term.}
\label{tab:tran_temp}
\end{table}

\subsection{Spectral function in a strongly magnetized medium within the ambit of effective models}
\begin{figure}[hbt]
\begin{center}
\subfigure[]
{\includegraphics[scale=0.35]{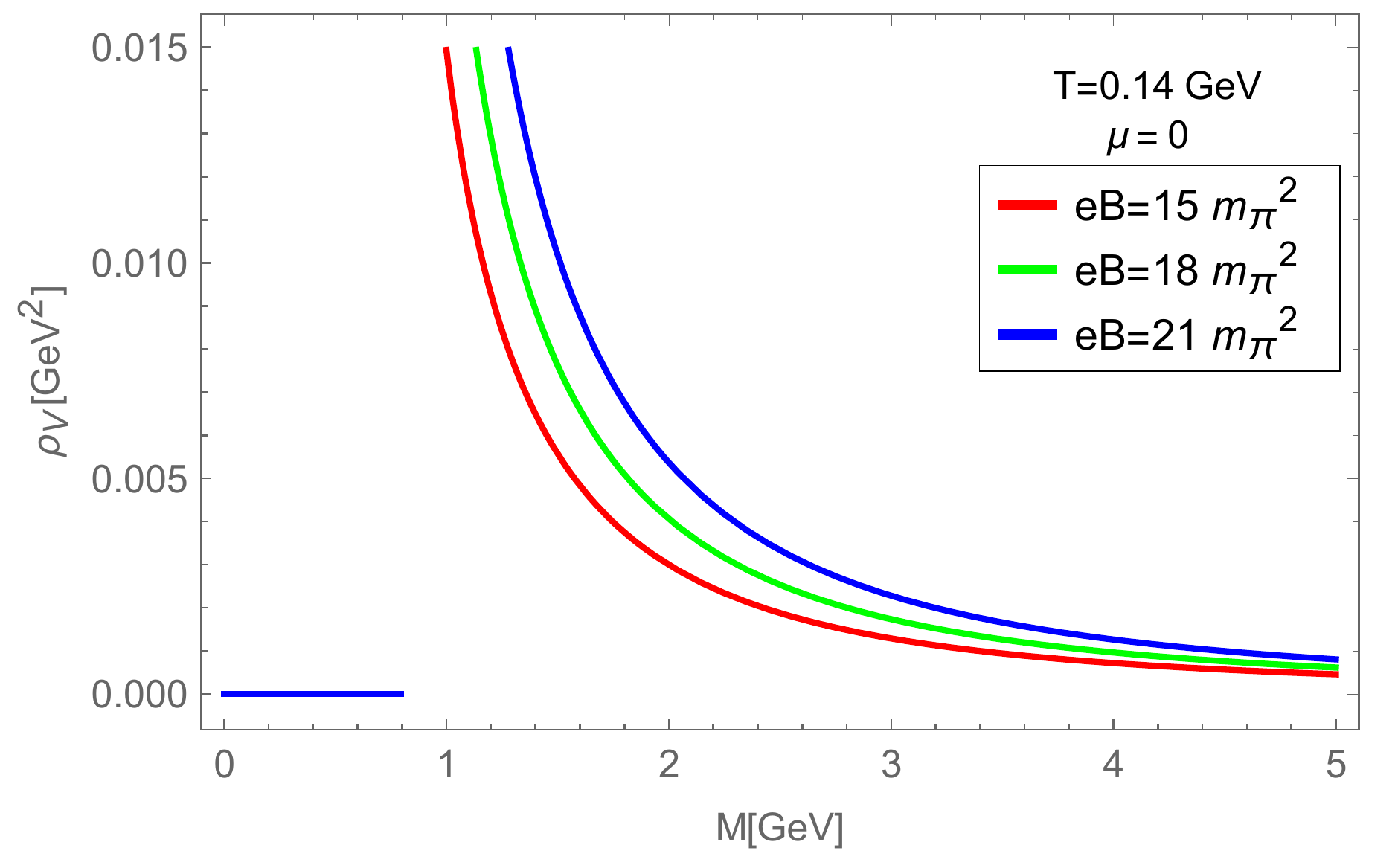}
\label{fig:spectral_func_vs_M_fixedT_njl}}
\subfigure[]
{\includegraphics[scale=0.35]{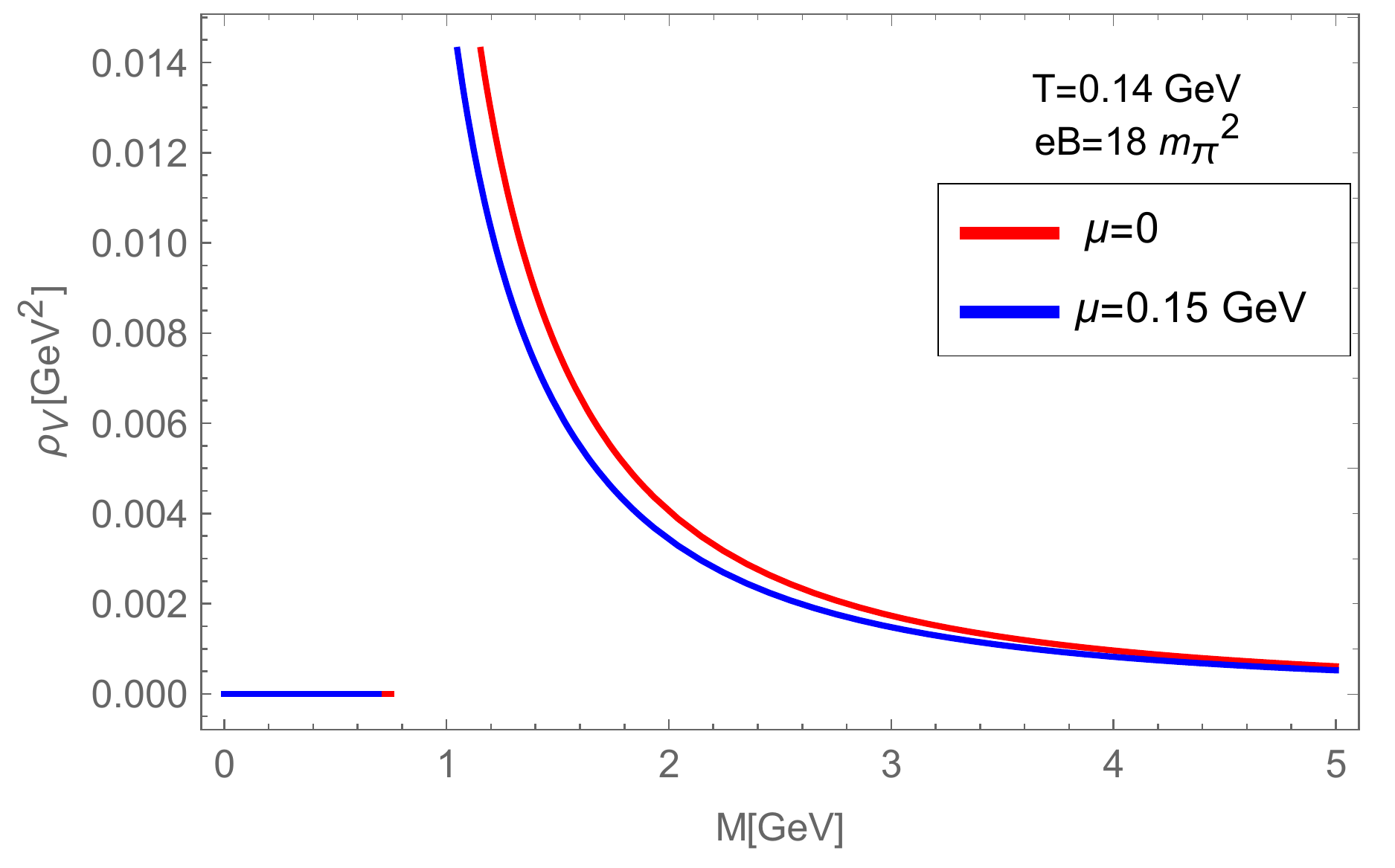}
\label{fig:spectral_func_vs_M_fixedT_nzmu_njl}}
\end{center}
\caption{Plot of the modified electromagnetic spectral function in a strongly magnetized medium with respect to the invariant mass of the external photon $M$ in NJL model. In the left panel the plot is done for $\mu=0$ and in the right we have comparison between zero and nonzero $\mu$.}
\label{fig:spectral_func_vs_M_fixedT}
\end{figure}

\begin{figure}[hbt]
\begin{center}
\subfigure[]
{\includegraphics[scale=0.35]{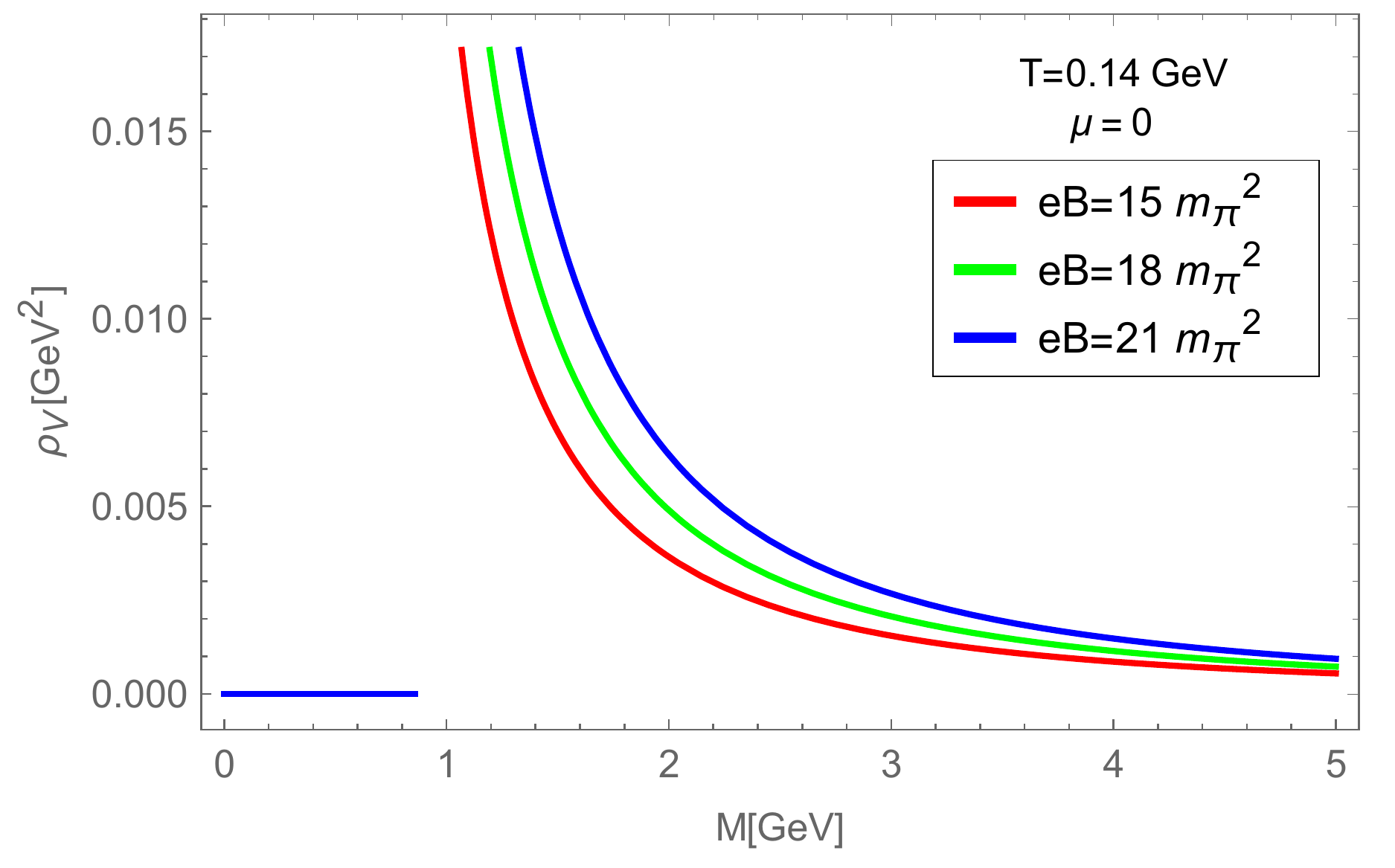}
\label{fig:spectral_func_vs_M_fixedT_pnjl}}
\subfigure[]
{\includegraphics[scale=0.35]{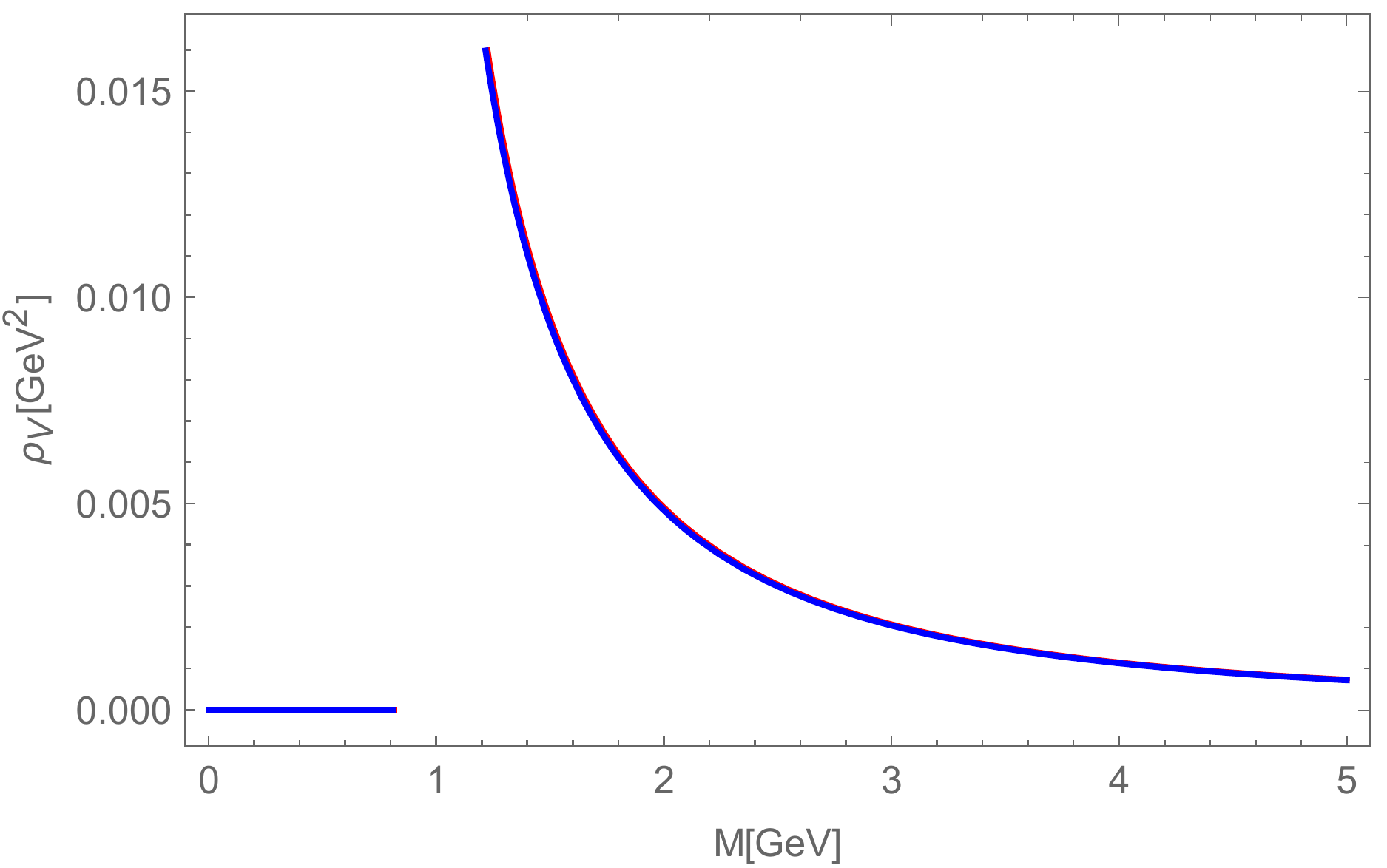}
\label{fig:spectral_func_vs_M_fixedT_nzmu_pnjl}}
\end{center}
\caption{Plot of the modified electromagnetic spectral function in a strongly magnetized medium with respect to the invariant mass of the external photon $M$ in PNJL model. In the left panel the plot is done for $\mu=0$ and in the right we have comparison between zero and nonzero $\mu$.}
\label{fig:spectral_func_vs_M_fixedT_pnjl_main}
\end{figure}

Next we move on to the main results of the present manuscript, i.e. the electromagnetic spectral properties in a strongly magnetized medium.  In figure~\ref{fig:spectral_func_vs_M_fixedT} the variation of the electromagnetic vector spectral function in NJL model is shown as a function of the invariant mass $M$ of the photon for different values of magnetic field with a fixed value of temperature and chemical potential. For the left panel~\ref{fig:spectral_func_vs_M_fixedT_njl} we have considered vanishing chemical potential, whereas in the right panel~\ref{fig:spectral_func_vs_M_fixedT_nzmu_njl} the effect of chemical potential on the spectral function has been shown for fixed values of both temperature and magnetic field. From fig~\ref{fig:spectral_func_vs_M_fixedT_njl} we learn that for given values of $T$ and $\mu$ the strength of the spectral function increases as we increase the magnetic field. Similarly it can be seen from figure~\ref{fig:spectral_func_vs_M_fixedT_nzmu_njl} that for a fixed value of $T$, with increased value of $\mu$ the strength of the spectral function have decreased. This is expected as the finite value of chemical potential adds to the medium effect.  Similar plots for PNJL model is shown in figure~\ref{fig:spectral_func_vs_M_fixedT_pnjl_main} where the effect of the chemical potential is seen to be negligible for the range of temperature that we have investigated for. Also one can notice the relative small increase in the value of the electromagnetic spectral function for the PNJL model as compared to the NJL one. This increment from NJL to PNJL can be even higher depending on the range of temperature that one deals with.

Now, since the spectral function calculation is done within the LLL approximation, the effective model inputs should comply with that assumption. As evident from figures~\ref{fig:M_njl},~\ref{fig:M_pnjl} and~\ref{fig:phivsT}, for models the LLL approximation is not valid for all temperatures. So depending on the value of temperature that we work with, it might introduce some error. For example, in figure~\ref{fig:spectral_func_vs_M_fixedT_njl} for the lowest considered magnetic field ($15\,m_\pi^2$), the LLL approximation introduces a maximum error of roughly $12\%$. The error gets reduced to $\sim7\%$ for $eB=21\,m_\pi^2$. These errors for NJL model can be further reduced by choosing appropriate temperature and chemical potential; for example the error gets reduced to below $5\%$ at $T=130$ MeV, $eB=21\,m_\pi^2$ and $\mu=0$ and it can be lowered by further reducing the temperature. The value of temperature for NJL model is so chosen that it can be compared with the PNJL one for which the maximum error from figure~\ref{fig:spectral_func_vs_M_fixedT_pnjl} is $\sim1\%$ for $eB=15\,m_\pi^2$ and becomes even lesser with the increase of magnetic field. Thus, for the same temperature the validity of LLL approximation is better in PNJL model as compared to the NJL one. This happens because of the presence of background gauge field which mimics the QCD more closely by emulating the confinement effect (statistically).

Finally if we compare our present results (in light of effective models) with the previous results obtained in Ref.~\cite{Bandyopadhyay:2016fyd} considering just the current quark mass, one can immediately notice a vast enhancement in the magnitude of the electromagnetic spectral function as we had expected \textendash\, it is roughly of the order of $10^4$. This can be understood in the following way. If we compare equation (26) from~\cite{Bandyopadhyay:2016fyd} with equation~(\ref{eqn:spec_sfa}) in this article, then it is clear that the major change is in terms of the mass. The effective mass is almost two orders of magnitude higher than the current quark mass that we used in the previous work. There are other changes also, for example the change in the distribution function, which is now dependent on the external magnetic field; but those are small as compared to the change in the value of quark mass.

\subsection{Dilepton production rate in a strongly magnetized medium within the ambit of effective models}
After the evaluation of the electromagnetic spectral functions in view of effective models, now we present our results for the dilepton production rate. As mentioned earlier, we will be dealing with two separate cases;  firstly when only the initial quarks are affected by the magnetic field, which can be applicable for a fast decaying external magnetic field or the dileptons produced at the edge of the fireball. And secondly the more general case of both the initial quarks and the produced lepton pairs being influenced by the external strong magnetic field. Below we discuss both the cases.

\subsubsection{Only the initial quarks are affected by the magnetic field}
\label{sssec:dr}
\begin{figure}[hbt]
\begin{center}
\subfigure[]
{\includegraphics[scale=0.35]{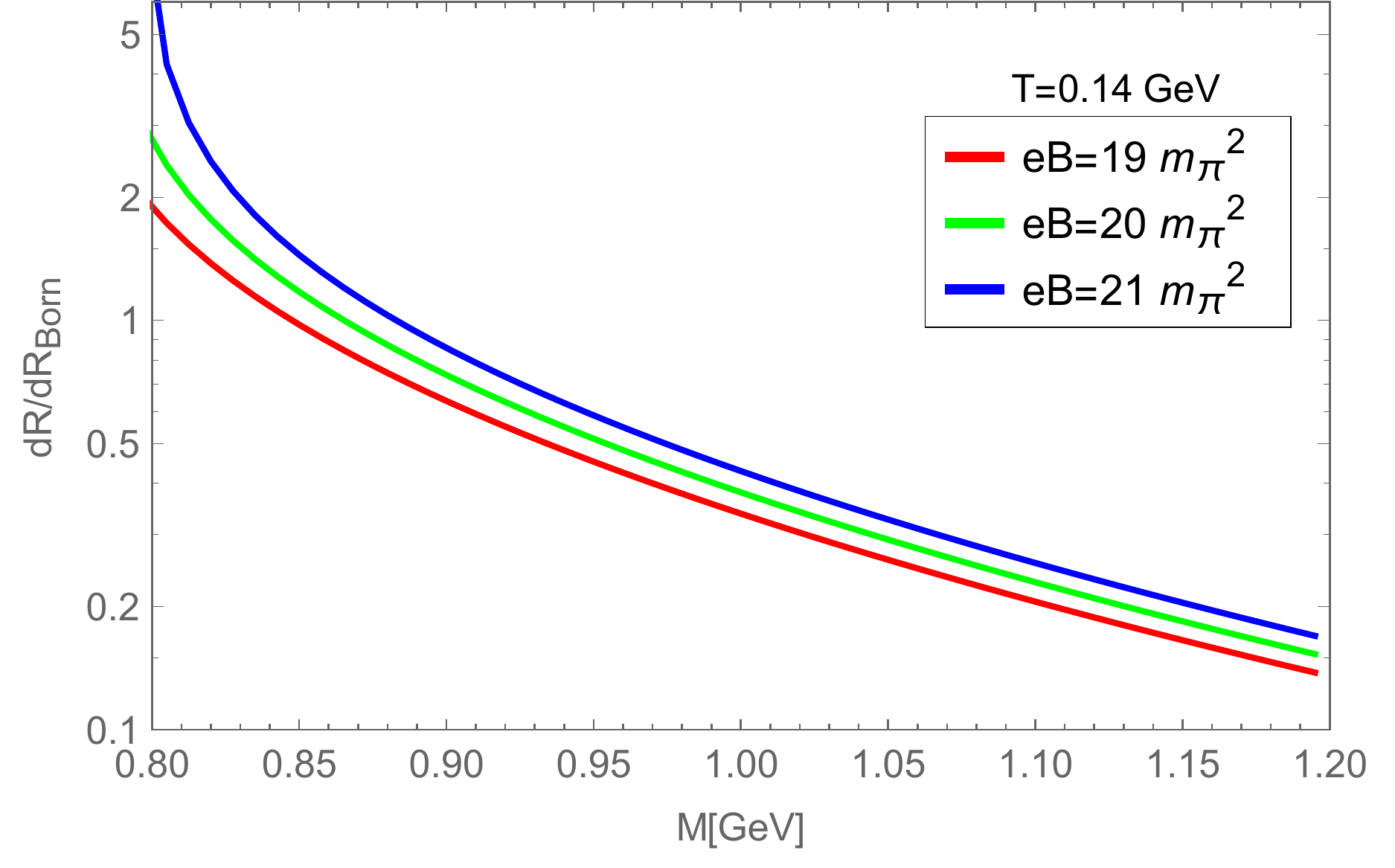}
\label{fig:dpr_njl_T140}}
\subfigure[]
{\includegraphics[scale=0.35]{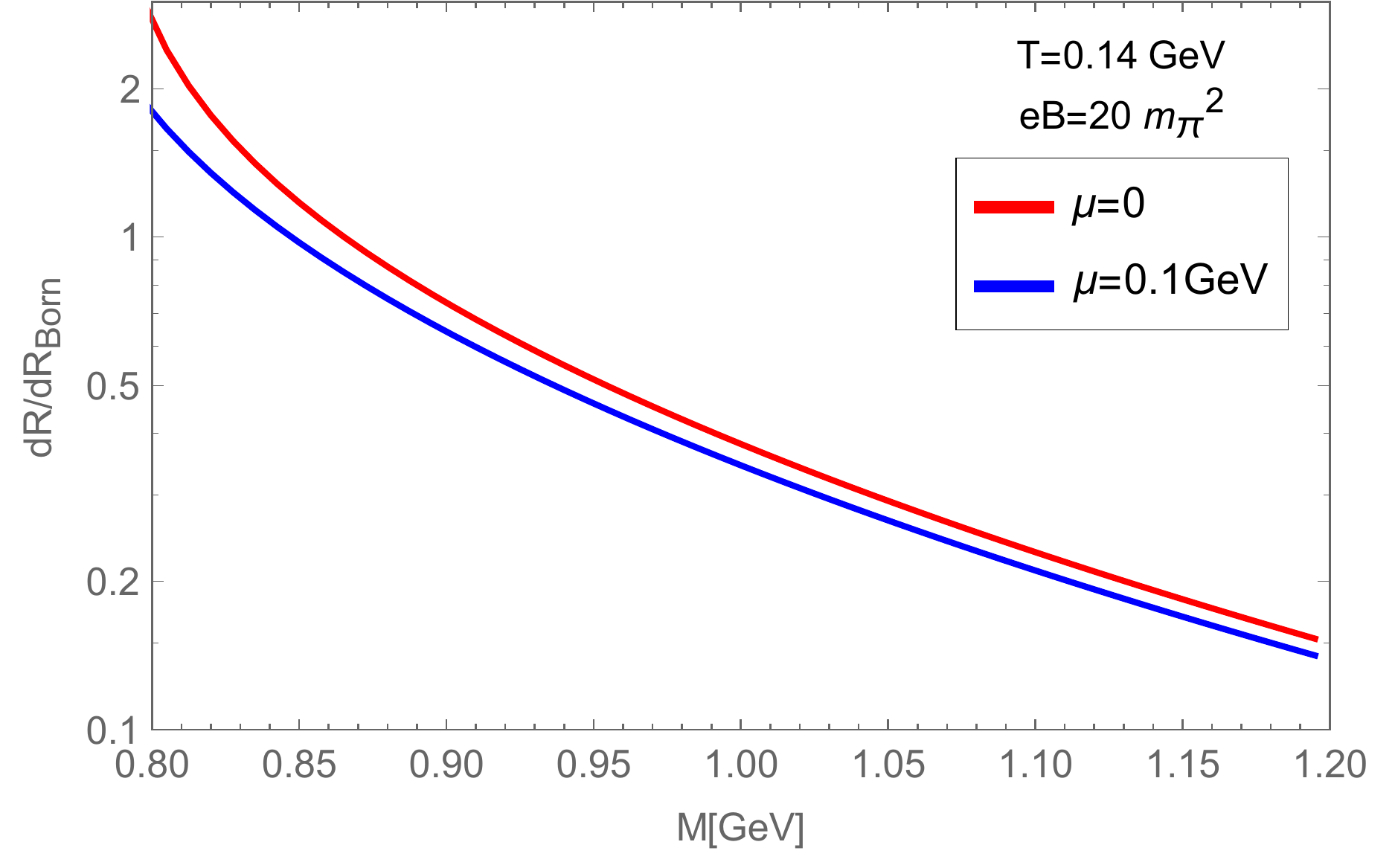}
\label{fig:dpr_njl_T140_nzmu}}
\end{center}
\caption{Plot of the ratio of dilepton production rate to Born rate in a strongly magnetized medium with respect to the invariant mass of the external photon $M$ in NJL model for zero chemical potential (left panel) and comparison with nonzero chemical potential (right panel) for different values of magnetic field and at a constant temperature kept fixed at 0.14 GeV. Only the initial quark pair is in the magnetic field.}
\label{fig:dpr_vs_M_fixedT_njl}
\end{figure}

\begin{figure}[hbt]
\begin{center}
\subfigure[]
{\includegraphics[scale=0.35]{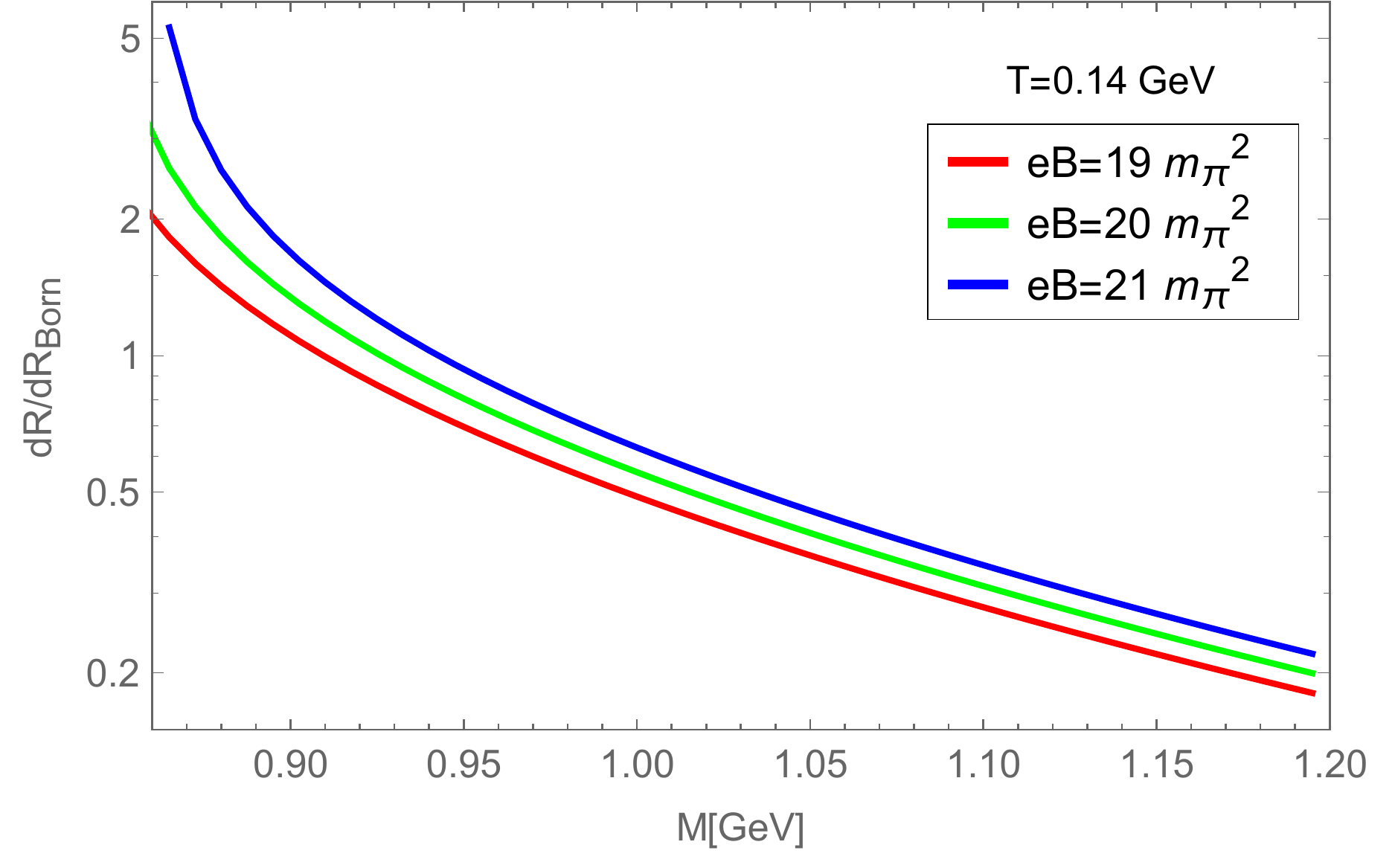}
\label{fig:dpr_pnjl_T140}}
\subfigure[]
{\includegraphics[scale=0.35]{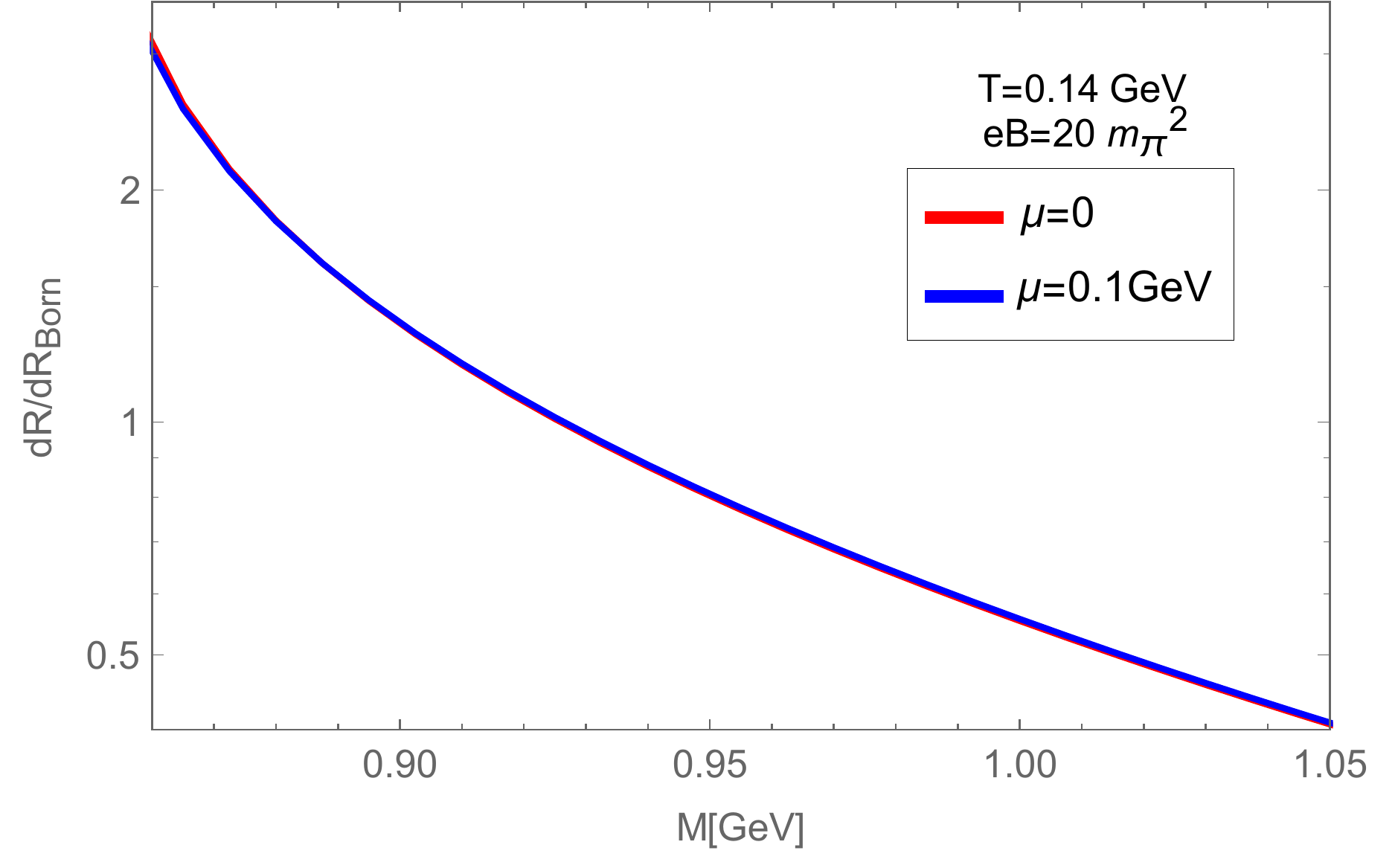}
\label{fig:dpr_pnjl_T140_nzmu}}
\end{center}
\caption{Plot of the ratio of dilepton production rate to Born rate in a strongly magnetized medium with respect to the invariant mass of the external photon $M$ in PNJL model for zero chemical potential (left panel) and comparison with nonzero chemical potential (right panel) for different values of magnetic field and at a constant temperature kept fixed at 0.14 GeV. Only the initial quark pair is in the magnetic field.}
\label{fig:dpr_vs_M_fixedT_pnjl}
\end{figure}
In figure~\ref{fig:dpr_vs_M_fixedT_njl}, the variation of the dilepton production rate in a strongly magnetized thermal medium scaled with the Born rate (i.e. Leading order dilepton production rate in a thermal medium) is shown as a function of the photon invariant mass for NJL model for different values of external magnetic fields at a fixed value of temperature. In the left panel~\ref{fig:dpr_njl_T140} we see that as we increase the strength of the magnetic field at a given value of invariant mass the rate also increases. In the right panel~\ref{fig:dpr_njl_T140_nzmu} the suppressing effect of chemical potential has been shown, which can be mapped from figure~\ref{fig:spectral_func_vs_M_fixedT_nzmu_njl}. Similar plots for PNJL are shown in figure~\ref{fig:dpr_vs_M_fixedT_pnjl}. The nature of all the plots can be easily understood extrapolating from the spectral function results. As an overall comparison with the existing results we found the following. In reference~\cite{Bandyopadhyay:2016fyd}, where this case was shown considering the current quark mass, the enhancement in the dilepton production rate (i.e. the ratio $dR/dR_{Born} > 1$ ) was for a very small range of invariant mass ($M<100$ MeV) which was insufficient to be detected in any of the heavy ion collisions experiment. Here for NJL model we can see that the enhancement is observed for a range of invariant mass starting at $M\sim800$ MeV and for PNJL one at $M\sim850$ MeV, which is certainly more compatible with the presently ongoing HIC experiments. The reason for such shifts in the values of invariant mass for which the DR is enhanced is the increase in effective mass which is dynamically generated in presence of magnetic field. After a certain value of invariant mass the DR falls off very fast because of LLL approximation~\cite{Bandyopadhyay:2016fyd}. While calculating the DR there is no more source of errors other than that already present in the spectral function. The maximum error possible in figure~\ref{fig:dpr_njl_T140} is $8\%$ while that in figure~\ref{fig:dpr_pnjl_T140} is less than $1\%$. For the temperature that we worked with the error in NJL model is larger than PNJL one. PNJL model is more relevant as far as the full QCD is concerned, thus the results in PNJL model should be taken more seriously.

\subsubsection{Both the initial quarks and the final lepton pairs are influenced by the magnetic field}
\begin{figure}[hbt]
\begin{center}
\subfigure[]
{\includegraphics[scale=0.35]{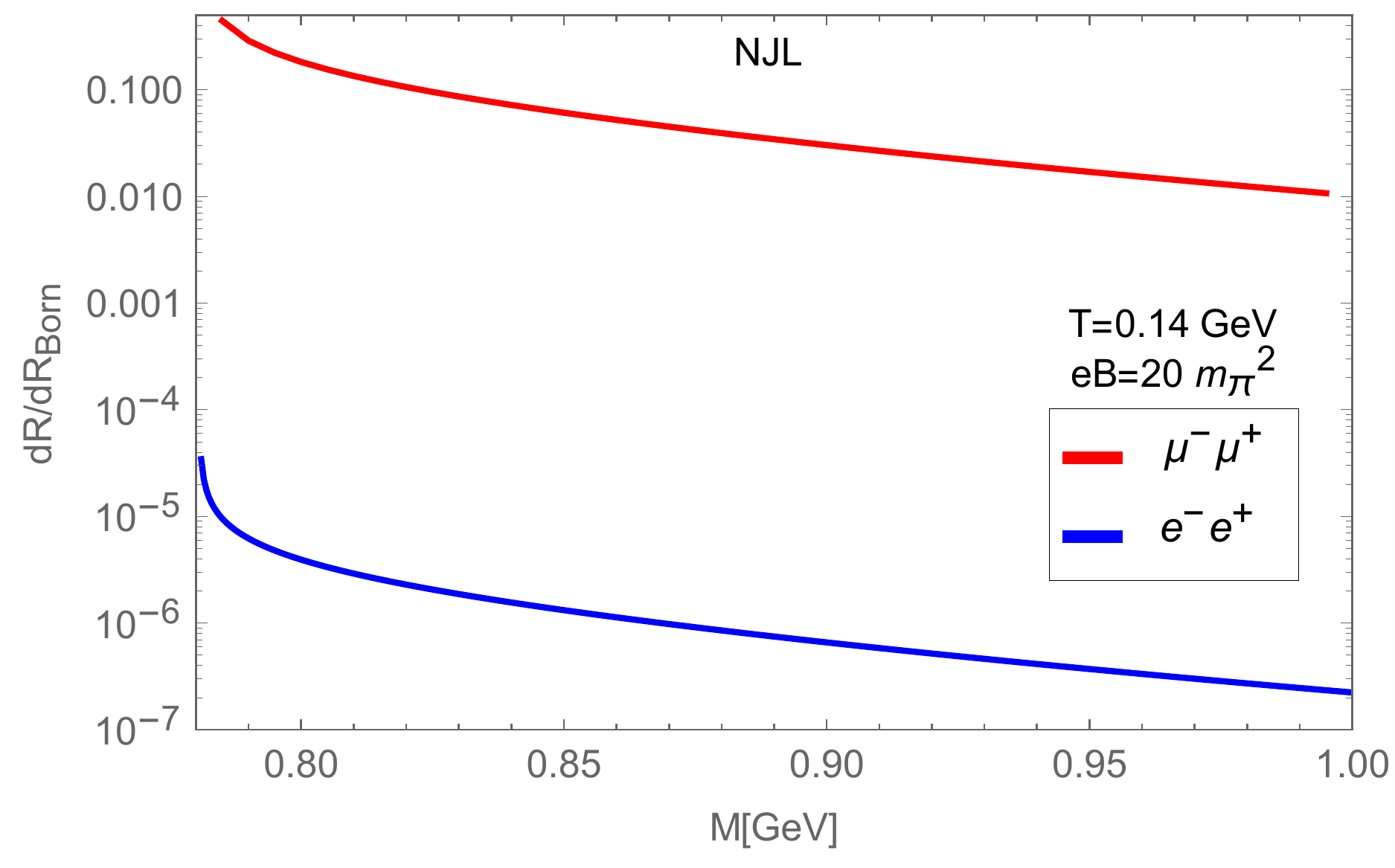}
\label{fig:dpr_njl_both_mag}}
\subfigure[]
{\includegraphics[scale=0.35]{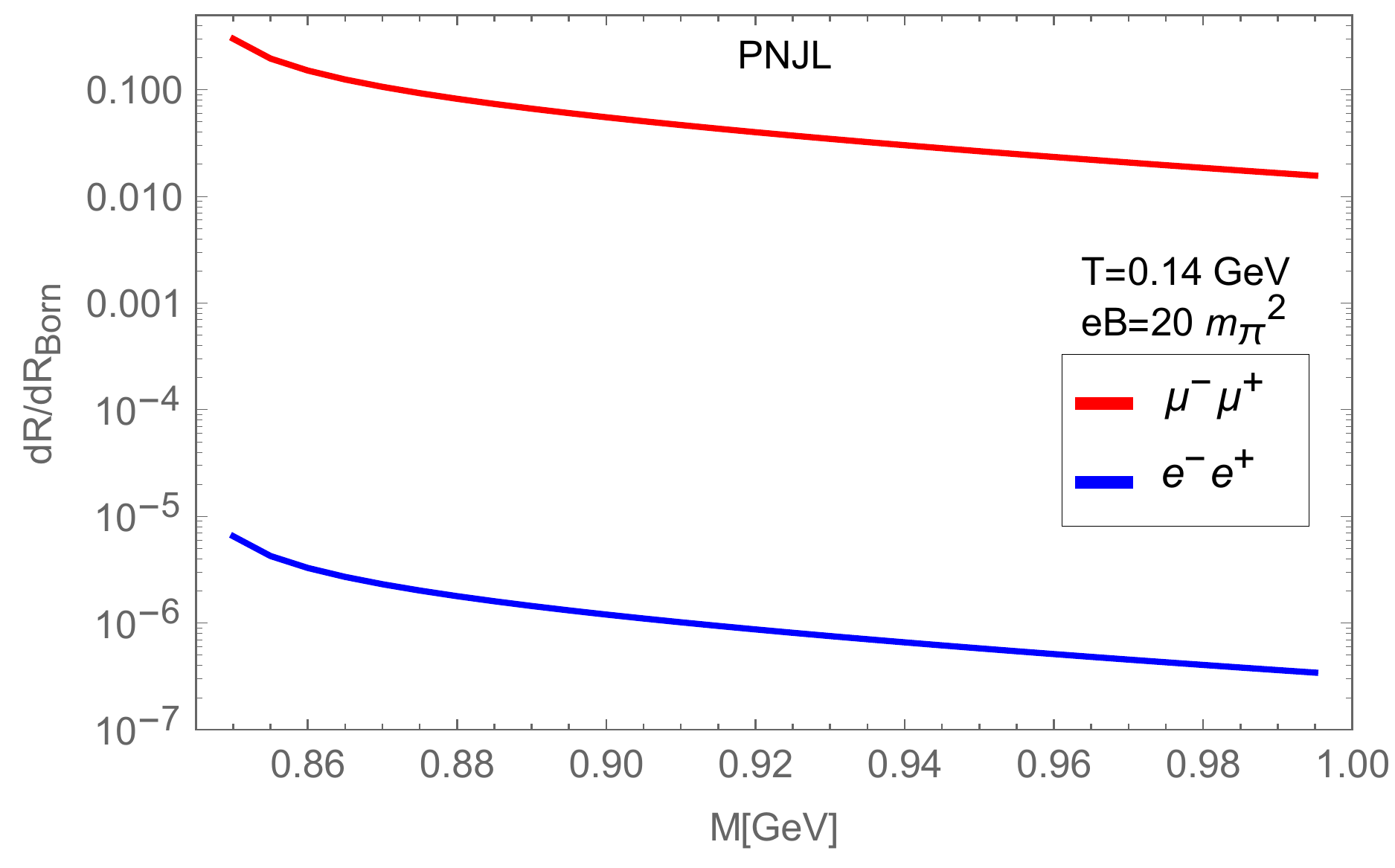}
\label{fig:dpr_pnjl_both_mag}}
\end{center}
\caption{Plot of the ratio of dilepton (both dimuon and dielectron) production rate to Born rate in a strongly magnetized medium with respect to the invariant mass of the external photon $M$ in NJL (left panel) and PNJL (right panel) models for $eB=20m_\pi^2$, $T=0.14$ GeV and $\mu=0$. Here both the quark pair and the final lepton pairs are affected by the magnetic field.}
\label{fig:dpr_vs_M_fixedT_both_mag}
\end{figure}
One would naively expect that the dynamical mass generation will also lead to the enhancement of DR when both initial quark pair and final lepton pair are influenced by the magnetic field. But this does not happen. In figure~\ref{fig:dpr_vs_M_fixedT_both_mag} we have shown the plots for both dimuon and dielectron in both NJL (\ref{fig:dpr_njl_both_mag}) and PNJL (\ref{fig:dpr_pnjl_both_mag}) models. First of all, in this case the DR is proportional to $B^2$ - where one $B$ comes from the leptonic part and the other from the electromagnetic spectral function involving quarks (vide equation~\ref{dcaseb}). Along with $M_f$ we have now another threshold in terms of the leptonic mass ($m_l$) and essentially the effective threshold is determined by ${\rm max}(M_f,m_l)$. We also observe that now the DR is proportional to $m_l^2$. The mass of the lepton is expected to be affected by the magnetic field, but we are in the strong field regime (LLL approximation) and the mass correction from the magnetic field vanishes. So we use the known vacuum values of masses for leptons. Because of the presence of these multiple factors which act opposite to $M_f$, the DR (ratio) in this case always remain below unity. The dimuon rate is $\sim 10^4$ times higher than the dielectron one, since the rate is proportional to $m_l^2$. The discussion on error calculation done in subsection~\ref{sssec:dr} remains valid for these results as well.

\section{Summary and Conclusion}
\label{sec:con}
In this work we have calculated the electromagnetic spectral function (SF) by evaluating the one-loop photon polarisation tensor for a strongly magnetised hot and dense medium. Because of the consideration of high enough magnetic field we use the lowest Landau level (LLL) approximation. The whole calculation is performed in the ambit of mean field models namely Nambu\textendash Jona-Lasinio (NJL) and its Polyakov loop extended version (PNJL). These models allow for a dynamical generation of a medium (i.e. $T$, $\mu$) dependent quark mass which further gets affected in the presence of a magnetised medium. This is shown here as magnetic catalysis (MC) for chiral transition in both NJL and PNJL models. The presence of magnetic field also influences the deconfinement dynamics \textendash \, a catalysing effect for deconfinement transition is also observed, although the effect is milder. We have also discussed both the importance and the effects of the Vandermonde (VdM) term in the pure gauge potential in presence of magnetic field. The value of the Polyakov loop, specially the form that we work with, overshoots unity for higher values of temperature in absence of the VdM term, which should not be the case. The inclusion of VdM term further shifts the critical temperature for both chiral and deconfinement transitions to higher values.

We have argued that for the values of the magnetic field considered here the LLL approximation is not strictly valid above certain respective temperatures. Keeping the temperature within the validity region of LLL we have evaluated the electromagnetic correlation function and dilepton rate (DR) using the dynamically generated magnetic field dependent effective mass of quark in the loop where LLL approximation is used. It is found that the strength of the SF gets boosted, predominantly because of the presence of dynamical quark mass. The difference in the orders of magnitude between the current quark mass and the dynamically generated effective mass gets reflected in the vast enhancement of the SF in our newfound results, when compared with the existing ones.  As we increase the strength of the magnetic field for a given value of temperature the strength of the SF increases, for both NJL and PNJL models. Being proportional to the SF this increment is further reflected in the DR which is enhanced as compared to the Born rate for certain values of invariant masses. For non-zero chemical potential the value decreases in NJL model due to increased opposing medium effect, whereas no significant change is observed in case of PNJL model for the range of temperatures that we have investigated for. Our most significant observation is the shifting of the enhanced DPR from an experimentally insignificant ($M<100$ MeV, in our previous work~\cite{Bandyopadhyay:2016fyd}) to an experimentally favourable range of invariant mass ($M \sim 800-900$ MeV), again mainly due to the magnetic field dependent dynamically generated effective mass.

It is important to mention here again that we have worked with the usual (P)NJL models which give rise to MC and we do not explicitly discuss the issues of IMC effects, mainly concentrating on the evaluation of the spectral function vis-a-vis dilepton rate.  This is reasonable as long as we are not close to the transition temperature, because away from the $T_c$ for both $T\ll T_c$ and $T\gg T_c$ one gets MC~\cite{Bali:2011qj,Bali:2012zg,Endrodi:2015oba}. It has also been established~\cite{Ferrer:2013noa} that introducing a new interaction term in the Lagrangian which preserves chiral symmetry and rotations along the magnetic field direction subsequently increases the effective dynamical mass in LLL. This would account for an exciting future prospect in view of spectral properties which, as we have shown in the present context, are heavily dependent on the dynamical mass. Some of the earlier studies~\cite{Frolov:2010wn, Ferrer:2019zfp} have also shown that in presence of baryonic chemical potential at zero or low temperatures, strong external magnetic field usually favors the formation of spatially non-uniform inhomogeneous condensates in the form of a dual chiral density waves (DCDW). As we have dealt with finite baryon chemical potential in this study, it would be interesting to see how the DCDW affects the spectral properties, both at zero and finite temperatures. Finally, we note that the LLL approximation valid for a high enough magnetic field may be realised only in limited cases in the present HIC experiments. To generalize our claim, the DR presented in this paper should be evaluated taking all the Landau levels into account, where the restriction on the temperature range may also be avoided. Work in this line is in progress.

\section{Acknowledgement}
\label{sec:ack}
CAI would like to thank Rishi Sharma for many fruitful discussions. He would also like to thank Department of Atomic Energy, India for the financial support. He acknowledges the facilities provided by Variable Energy Cyclotron Centre, India where he was a visiting fellow when the preliminary parts of the work were completed and also his present institute, Tata Institute of Fundamental Research, India.  AB is supported by Conselho Nacional de Desenvolvimento Cient\'{i}fico e Tecnol\'{o}gico (CNPq), Govt. of Brazil, under grant CAPES.


\begin{thebibliography}{99}

\bibitem{Kharzeev:2013jha} 
  D.~Kharzeev, K.~Landsteiner, A.~Schmitt and H.~U.~Yee,
  Lect.\ Notes Phys.\  {\bf 871}, pp.1 (2013).
  doi:10.1007/978-3-642-37305-3
  
\bibitem{Lattimer:2006xb} 
  J.~M.~Lattimer and M.~Prakash,
  Phys.\ Rept.\  {\bf 442}, 109 (2007)
  doi:10.1016/j.physrep.2007.02.003
  [astro-ph/0612440].
  
\bibitem{Ferrer:2005vd} 
  E.~J.~Ferrer, V.~de la Incera and C.~Manuel,
  Phys.\ Rev.\ Lett.\  {\bf 95}, 152002 (2005)
  doi:10.1103/PhysRevLett.95.152002
  [hep-ph/0503162].  
  
\bibitem{Ferrer:2006vw} 
  E.~J.~Ferrer, V.~de la Incera and C.~Manuel,
  Nucl.\ Phys.\ B {\bf 747}, 88 (2006)
  doi:10.1016/j.nuclphysb.2006.04.013
  [hep-ph/0603233].  
  
\bibitem{Ferrer:2007iw} 
  E.~J.~Ferrer and V.~de la Incera,
  Phys.\ Rev.\ D {\bf 76}, 045011 (2007)
  doi:10.1103/PhysRevD.76.045011
  [nucl-th/0703034 [NUCL-TH]].  

\bibitem{Fukushima:2007fc} 
  K.~Fukushima and H.~J.~Warringa,
  Phys.\ Rev.\ Lett.\  {\bf 100}, 032007 (2008)
  doi:10.1103/PhysRevLett.100.032007
  [arXiv:0707.3785 [hep-ph]].
  
\bibitem{Noronha:2007wg} 
  J.~L.~Noronha and I.~A.~Shovkovy,
  Phys.\ Rev.\ D {\bf 76}, 105030 (2007)
  Erratum: [Phys.\ Rev.\ D {\bf 86}, 049901 (2012)]
  doi:10.1103/PhysRevD.76.105030, 10.1103/PhysRevD.86.049901
  [arXiv:0708.0307 [hep-ph]].  
  
\bibitem{Feng:2009vt} 
  B.~Feng, D.~Hou, H.~c.~Ren and P.~p.~Wu,
  Phys.\ Rev.\ Lett.\  {\bf 105}, 042001 (2010)
  doi:10.1103/PhysRevLett.105.042001
  [arXiv:0911.4997 [hep-ph]].  
  
\bibitem{Skokov:2009qp} 
  V.~Skokov, A.~Y.~Illarionov and V.~Toneev,
  Int.\ J.\ Mod.\ Phys.\ A {\bf 24}, 5925 (2009)
  doi:10.1142/S0217751X09047570
  [arXiv:0907.1396 [nucl-th]].  
  
\bibitem{Vachaspati:1991nm} 
  T.~Vachaspati,
  Phys.\ Lett.\ B {\bf 265}, 258 (1991).
  doi:10.1016/0370-2693(91)90051-Q  
  
\bibitem{Muller:1983ed} 
  B.~Muller,
  ``The Physics Of The Quark - Gluon Plasma,''
  Lect.\ Notes Phys.\  {\bf 225}, 1 (1985).
  
\bibitem{Heinz:2000bk} 
  U.~W.~Heinz and M.~Jacob,
  ``Evidence for a new state of matter: An Assessment of the results from the CERN lead beam program,''
  nucl-th/0002042.
  
\bibitem{Alexandre:2000yf} 
  J.~Alexandre, K.~Farakos and G.~Koutsoumbas,
  Phys.\ Rev.\ D {\bf 63}, 065015 (2001)
  doi:10.1103/PhysRevD.63.065015
  [hep-th/0010211].

\bibitem{Gusynin:1997kj} 
  V.~P.~Gusynin and I.~A.~Shovkovy,
  Phys.\ Rev.\ D {\bf 56}, 5251 (1997)
  doi:10.1103/PhysRevD.56.5251
  [hep-ph/9704394].

\bibitem{Lee:1997zj} 
  D.~S.~Lee, C.~N.~Leung and Y.~J.~Ng,
  Phys.\ Rev.\ D {\bf 55}, 6504 (1997)
  doi:10.1103/PhysRevD.55.6504
  [hep-th/9701172].

\bibitem{Bali:2011qj} 
  G.~S.~Bali, F.~Bruckmann, G.~Endrodi, Z.~Fodor, S.~D.~Katz, S.~Krieg, A.~Schafer and K.~K.~Szabo,
  JHEP {\bf 1202}, 044 (2012)
  doi:10.1007/JHEP02(2012)044
  [arXiv:1111.4956 [hep-lat]].

\bibitem{Bornyakov:2013eya} 
  V.~G.~Bornyakov, P.~V.~Buividovich, N.~Cundy, O.~A.~Kochetkov and A.~Schäfer,
  Phys.\ Rev.\ D {\bf 90}, no. 3, 034501 (2014)
  doi:10.1103/PhysRevD.90.034501
  [arXiv:1312.5628 [hep-lat]].
 
\bibitem{Mueller:2015fka} 
  N.~Mueller and J.~M.~Pawlowski,
  Phys.\ Rev.\ D {\bf 91}, no. 11, 116010 (2015)
  doi:10.1103/PhysRevD.91.116010
  [arXiv:1502.08011 [hep-ph]].
 
\bibitem{Ayala:2014iba} 
  A.~Ayala, M.~Loewe, A.~J.~Mizher and R.~Zamora,
  Phys.\ Rev.\ D {\bf 90}, no. 3, 036001 (2014)
  doi:10.1103/PhysRevD.90.036001
  [arXiv:1406.3885 [hep-ph]].
 
\bibitem{Ayala:2014gwa} 
  A.~Ayala, M.~Loewe and R.~Zamora,
  Phys.\ Rev.\ D {\bf 91}, no. 1, 016002 (2015)
  doi:10.1103/PhysRevD.91.016002
  [arXiv:1406.7408 [hep-ph]].
  
\bibitem{Ayala:2016sln} 
  A.~Ayala, M.~Loewe and R.~Zamora,
  J.\ Phys.\ Conf.\ Ser.\  {\bf 720}, 012026 (2016).
  
\bibitem{Ayala:2015bgv} 
  A.~Ayala, C.~A.~Dominguez, L.~A.~Hernandez, M.~Loewe and R.~Zamora,
  Phys.\ Lett.\ B {\bf 759}, 99 (2016),
  [arXiv:1510.09134 [hep-ph]].
  
\bibitem{Kharzeev:2007jp} 
  D.~E.~Kharzeev, L.~D.~McLerran and H.~J.~Warringa,
  Nucl.\ Phys.\ A {\bf 803}, 227 (2008)
  doi:10.1016/j.nuclphysa.2008.02.298
  [arXiv:0711.0950 [hep-ph]].
 
\bibitem{Fukushima:2008xe} 
  K.~Fukushima, D.~E.~Kharzeev and H.~J.~Warringa,
  Phys.\ Rev.\ D {\bf 78}, 074033 (2008)
  doi:10.1103/PhysRevD.78.074033
  [arXiv:0808.3382 [hep-ph]].
 
\bibitem{Kharzeev:2009fn} 
  D.~E.~Kharzeev,
  Annals Phys.\  {\bf 325}, 205 (2010)
  doi:10.1016/j.aop.2009.11.002
  [arXiv:0911.3715 [hep-ph]].

\bibitem{Fayazbakhsh:2010bh} 
  S.~Fayazbakhsh and N.~Sadooghi,
  Phys.\ Rev.\ D {\bf 83}, 025026 (2011)
  doi:10.1103/PhysRevD.83.025026
  [arXiv:1009.6125 [hep-ph]].
 
\bibitem{Fayazbakhsh:2010gc} 
  S.~Fayazbakhsh and N.~Sadooghi,
  Phys.\ Rev.\ D {\bf 82}, 045010 (2010)
  doi:10.1103/PhysRevD.82.045010
  [arXiv:1005.5022 [hep-ph]].
 
\bibitem{Sadooghi:2015hha} 
  N.~Sadooghi and F.~Taghinavaz,
  Phys.\ Rev.\ D {\bf 92}, no. 2, 025006 (2015)
  doi:10.1103/PhysRevD.92.025006
  [arXiv:1504.04268 [hep-ph]].
 
\bibitem{Das:2017vfh} 
  A.~Das, A.~Bandyopadhyay, P.~K.~Roy and M.~G.~Mustafa,
  Phys.\ Rev.\ D {\bf 97}, no. 3, 034024 (2018)
  doi:10.1103/PhysRevD.97.034024
  [arXiv:1709.08365 [hep-ph]]. 
  
\bibitem{Karmakar:2018aig} 
  B.~Karmakar, A.~Bandyopadhyay, N.~Haque and M.~G.~Mustafa,
  arXiv:1804.11336 [hep-ph].  
 
\bibitem{Strickland:2012vu} 
  M.~Strickland, V.~Dexheimer and D.~P.~Menezes,
  Phys.\ Rev.\ D {\bf 86}, 125032 (2012)
  doi:10.1103/PhysRevD.86.125032
  [arXiv:1209.3276 [nucl-th]].
 
\bibitem{Andersen:2014xxa} 
  J.~O.~Andersen, W.~R.~Naylor and A.~Tranberg,
  Rev.\ Mod.\ Phys.\  {\bf 88}, 025001 (2016)
  doi:10.1103/RevModPhys.88.025001
  [arXiv:1411.7176 [hep-ph]].

\bibitem{Bandyopadhyay:2017cle} 
  A.~Bandyopadhyay, B.~Karmakar, N.~Haque and M.~G.~Mustafa,
  arXiv:1702.02875 [hep-ph].
  
\bibitem{Tuchin:2012mf} 
  K.~Tuchin,
  Phys.\ Rev.\ C {\bf 87}, no. 2, 024912 (2013)
  doi:10.1103/PhysRevC.87.024912
  [arXiv:1206.0485 [hep-ph]].

\bibitem{Tuchin:2013bda} 
  K.~Tuchin,
  Phys.\ Rev.\ C {\bf 88}, 024910 (2013)
  doi:10.1103/PhysRevC.88.024910
  [arXiv:1305.0545 [nucl-th]].
  
\bibitem{Tuchin:2013ie} 
  K.~Tuchin,
  Adv.\ High Energy Phys.\  {\bf 2013}, 490495 (2013)
  doi:10.1155/2013/490495
  [arXiv:1301.0099 [hep-ph]].

\bibitem{Sadooghi:2016jyf} 
  N.~Sadooghi and F.~Taghinavaz,
  Annals Phys.\  {\bf 376}, 218 (2017)
  doi:10.1016/j.aop.2016.11.008
  [arXiv:1601.04887 [hep-ph]].
   
\bibitem{Bandyopadhyay:2016fyd} 
  A.~Bandyopadhyay, C.~A.~Islam and M.~G.~Mustafa,
  Phys.\ Rev.\ D {\bf 94}, no. 11, 114034 (2016)
  doi:10.1103/PhysRevD.94.114034
  [arXiv:1602.06769 [hep-ph]].
  
\bibitem{Bandyopadhyay:2017raf} 
  A.~Bandyopadhyay and S.~Mallik,
  Phys.\ Rev.\ D {\bf 95}, no. 7, 074019 (2017)
  doi:10.1103/PhysRevD.95.074019
  [arXiv:1704.01364 [hep-ph]].  
  
\bibitem{Ghosh:2018xhh} 
  S.~Ghosh and V.~Chandra,
  Phys.\ Rev.\ D {\bf 98}, no. 7, 076006 (2018)
  doi:10.1103/PhysRevD.98.076006
  [arXiv:1808.05176 [hep-ph]].  
  
\bibitem{Forster(book):1975HFBSCF} D. Forster, Hydrodynamics Fluctuation, Broken Symmetry and Correlation Function 
(Benjamin/Cummings, Menlo Park,CA, 1975);

\bibitem{Callen:1951vq} 
  H.~B.~Callen and T.~A.~Welton,
  ``Irreversibility and generalized noise,''
  Phys.\ Rev.\  {\bf 83}, 34 (1951).
  
\bibitem{Kubo:1957mj} 
  R.~Kubo,
  ``Statistical mechanical theory of irreversible processes. 1. General theory and simple applications 
  in magnetic and conduction problems,''
  J.\ Phys.\ Soc.\ Jap.\  {\bf 12}, 570 (1957).  
  
\bibitem{Davidson:1995fq} 
  R.~M.~Davidson and E.~Ruiz Arriola,
  ``Mesonic correlation functions in the NJL model with vector mesons,''
  Phys.\ Lett.\ B {\bf 359}, 273 (1995).  
  
\bibitem{Kapusta_Gale(book):1996FTFTPA} J. I. Kapusta and C. Gale, Finite Temperature Field 
Theory Principle and Applications (Cambridge University Press,
Cambridge, 1996), 2nd ed.

\bibitem{Lebellac(book):1996TFT} M. LeBellac, Thermal Field Theory 
(Cambridge University Press, Cambridge, 1996), 1st ed.
  
\bibitem{Jackson(book):1975CE} J. Jackson, \textit {Classical Electrodynamics}, 
2nd ed. ~Wiley, New
York, 1975; R. Dalitz and D. Yennie, Phys. Rev. {\bf 105}, 1598  (1957).

\bibitem{Schwinger:1951nm} 
  J.~S.~Schwinger,
  Phys.\ Rev.\  {\bf 82}, 664 (1951).
  doi:10.1103/PhysRev.82.664
  
\bibitem{Islam:2014sea} 
  C.~A.~Islam, S.~Majumder, N.~Haque and M.~G.~Mustafa,
  JHEP {\bf 1502}, 011 (2015)
  doi:10.1007/JHEP02(2015)011
  [arXiv:1411.6407 [hep-ph]].   
  
\bibitem{Nambu:1961tp} 
  Y.~Nambu and G.~Jona-Lasinio,
  Phys.\ Rev.\  {\bf 122}, 345 (1961).
  doi:10.1103/PhysRev.122.345  
  
\bibitem{Nambu:1961fr} 
  Y.~Nambu and G.~Jona-Lasinio,
  Phys.\ Rev.\  {\bf 124}, 246 (1961).
  doi:10.1103/PhysRev.124.246  
  
\bibitem{Klevansky:1992qe} 
  S.~P.~Klevansky,
  ``The Nambu-Jona-Lasinio model of quantum chromodynamics,''
  Rev.\ Mod.\ Phys.\  {\bf 64}, 649 (1992).  
  
\bibitem{Fukushima:2003fw} 
  K.~Fukushima,
  ``Chiral effective model with the Polyakov loop,''
  Phys.\ Lett.\ B {\bf 591}, 277 (2004)
  [hep-ph/0310121].
  
\bibitem{Fukushima:2003fm} 
  K.~Fukushima,
  ``Relation between the Polyakov loop and the chiral order parameter at strong coupling,''
  Phys.\ Rev.\ D {\bf 68}, 045004 (2003)
  [hep-ph/0303225].
  
\bibitem{Ratti:2005jh} 
  C.~Ratti, M.~A.~Thaler and W.~Weise,
  ``Phases of QCD: Lattice thermodynamics and a field theoretical model,''
  Phys.\ Rev.\ D {\bf 73}, 014019 (2006)
  [hep-ph/0506234].  
  
\bibitem{Menezes:2008qt} 
  D.~P.~Menezes, M.~Benghi Pinto, S.~S.~Avancini, A.~Perez Martinez and C.~Providencia,
  Phys.\ Rev.\ C {\bf 79}, 035807 (2009)
  doi:10.1103/PhysRevC.79.035807
  [arXiv:0811.3361 [nucl-th]].  
  
\bibitem{Boomsma:2009yk} 
  J.~K.~Boomsma and D.~Boer,
  Phys.\ Rev.\ D {\bf 81}, 074005 (2010)
  doi:10.1103/PhysRevD.81.074005
  [arXiv:0911.2164 [hep-ph]].  
  
\bibitem{Chatterjee:2011ry} 
  B.~Chatterjee, H.~Mishra and A.~Mishra,
  Phys.\ Rev.\ D {\bf 84}, 014016 (2011)
  doi:10.1103/PhysRevD.84.014016
  [arXiv:1101.0498 [hep-ph]].  
  
\bibitem{Avancini:2011zz} 
  S.~S.~Avancini, D.~P.~Menezes and C.~Providencia,
  Phys.\ Rev.\ C {\bf 83}, 065805 (2011).
  doi:10.1103/PhysRevC.83.065805  
  
\bibitem{Farias:2014eca} 
  R.~L.~S.~Farias, K.~P.~Gomes, G.~I.~Krein and M.~B.~Pinto,
  Phys.\ Rev.\ C {\bf 90}, no. 2, 025203 (2014)
  doi:10.1103/PhysRevC.90.025203
  [arXiv:1404.3931 [hep-ph]].  
  
\bibitem{Ferrer:2014qka} 
  E.~J.~Ferrer, V.~de la Incera and X.~J.~Wen,
  Phys.\ Rev.\ D {\bf 91}, no. 5, 054006 (2015)
  doi:10.1103/PhysRevD.91.054006
  [arXiv:1407.3503 [nucl-th]].  
  
\bibitem{Yu:2014xoa} 
  L.~Yu, J.~Van Doorsselaere and M.~Huang,
  Phys.\ Rev.\ D {\bf 91}, no. 7, 074011 (2015)
  doi:10.1103/PhysRevD.91.074011
  [arXiv:1411.7552 [hep-ph]].  

\bibitem{Mao:2016fha} 
  S.~Mao,
  Phys.\ Lett.\ B {\bf 758}, 195 (2016)
  doi:10.1016/j.physletb.2016.05.018
  [arXiv:1602.06503 [hep-ph]].
  
\bibitem{Farias:2016gmy} 
  R.~L.~S.~Farias, V.~S.~Timoteo, S.~S.~Avancini, M.~B.~Pinto and G.~Krein,
  Eur.\ Phys.\ J.\ A {\bf 53}, no. 5, 101 (2017)
  doi:10.1140/epja/i2017-12320-8
  [arXiv:1603.03847 [hep-ph]].  
  
\bibitem{Adler:2006yt} 
  S.~S.~Adler {\it et al.}  [PHENIX Collaboration],
  ``Measurement of direct photon production in p + p collisions at $\sqrt {s} = 200$-GeV,''
  Phys.\ Rev.\ Lett.\  {\bf 98}, 012002 (2007)
  [hep-ex/0609031].
  
\bibitem{Adare:2006ti} 
  A.~Adare {\it et al.}  [PHENIX Collaboration],
  ``Scaling properties of azimuthal anisotropy in Au+Au and Cu+Cu collisions at $\sqrt{s_{NN}}$ = 200-GeV,''
  Phys.\ Rev.\ Lett.\  {\bf 98}, 162301 (2007)
  [nucl-ex/0608033].
  
\bibitem{Adare:2006nq} 
  A.~Adare {\it et al.}  [PHENIX Collaboration],
  ``Energy Loss and Flow of Heavy Quarks in Au+Au Collisions at $\sqrt {s_{NN}}= 200$-GeV,''
  Phys.\ Rev.\ Lett.\  {\bf 98}, 172301 (2007)
  [nucl-ex/0611018].
  
\bibitem{Abelev:2006db} 
  B.~I.~Abelev {\it et al.}  [STAR Collaboration],
  ``Erratum: Transverse momentum and centrality dependence of high-$p_T$ non-photonic 
  electron suppression in Au+Au collisions at $\sqrt{s_{NN}} = 200$\,GeV,''
  Phys.\ Rev.\ Lett.\  {\bf 98}, 192301 (2007)
  [Erratum-ibid.\  {\bf 106}, 159902 (2011)]
  [nucl-ex/0607012].
  
\bibitem{Aamodt:2010pb} 
  K.~Aamodt {\it et al.}  [ALICE Collaboration],
  ``Charged-particle multiplicity density at mid-rapidity in central Pb-Pb collisions at $\sqrt{s_{NN}} = 2.76$ TeV,''
  Phys.\ Rev.\ Lett.\  {\bf 105}, 252301 (2010)
  [arXiv:1011.3916 [nucl-ex]].

\bibitem{Aamodt:2010pa} 
  K.~Aamodt {\it et al.}  [ALICE Collaboration],
  ``Elliptic flow of charged particles in Pb-Pb collisions at 2.76 TeV,''
  Phys.\ Rev.\ Lett.\  {\bf 105}, 252302 (2010)
  [arXiv:1011.3914 [nucl-ex]].
  
  \bibitem{Aamodt:2010jd} 
  K.~Aamodt {\it et al.}  [ALICE Collaboration],
  ``Suppression of Charged Particle Production at Large Transverse Momentum in Central Pb--Pb Collisions at $\sqrt{s_{NN}} = 2.76$ TeV,''
  Phys.\ Lett.\ B {\bf 696}, 30 (2011)
  [arXiv:1012.1004 [nucl-ex]].

\bibitem{Aad:2010bu} 
  G.~Aad {\it et al.}  [ATLAS Collaboration],
  ``Observation of a Centrality-Dependent Dijet Asymmetry in Lead-Lead Collisions at $\sqrt{s_{NN}}=2.77$ TeV with the ATLAS Detector at the LHC,''
  Phys.\ Rev.\ Lett.\  {\bf 105}, 252303 (2010)
  [arXiv:1011.6182 [hep-ex]].  
  
\bibitem{Gatto:2010qs}
  R.~Gatto and M.~Ruggieri,
  Phys.\ Rev.\ D {\bf 82} (2010) 054027
  doi:10.1103/PhysRevD.82.054027
  [arXiv:1007.0790 [hep-ph]].  
  
\bibitem{Providencia:2014txa} 
  C.~Providência, M.~Ferreira and P.~Costa,
  Acta Phys.\ Polon.\ Supp.\  {\bf 8}, no. 1, 207 (2015)
  doi:10.5506/APhysPolBSupp.8.207
  [arXiv:1412.8308 [hep-ph]]. 
  
\bibitem{Ferreira:2013tba} 
  M.~Ferreira, P.~Costa, D.~P.~Menezes, C.~Providência and N.~Scoccola,
  Phys.\ Rev.\ D {\bf 89}, no. 1, 016002 (2014)
  Addendum: [Phys.\ Rev.\ D {\bf 89}, no. 1, 019902 (2014)]
  doi:10.1103/PhysRevD.89.016002, 10.1103/PhysRevD.89.019902
  [arXiv:1305.4751 [hep-ph]].  
  
\bibitem{Ferreira:2014kpa} 
  M.~Ferreira, P.~Costa, O.~Lourenço, T.~Frederico and C.~Providência,
  Phys.\ Rev.\ D {\bf 89}, no. 11, 116011 (2014)
  doi:10.1103/PhysRevD.89.116011
  [arXiv:1404.5577 [hep-ph]].
  
\bibitem{Buballa:2003qv} 
  M.~Buballa,
  Phys.\ Rept.\  {\bf 407}, 205 (2005)
  doi:10.1016/j.physrep.2004.11.004
  [hep-ph/0402234].
  
\bibitem{Mukherjee:2006hq} 
  S.~Mukherjee, M.~G.~Mustafa and R.~Ray,
  ``Thermodynamics of the PNJL model with nonzero baryon and isospin chemical potentials,''
  Phys.\ Rev.\ D {\bf 75}, 094015 (2007)
  [hep-ph/0609249].

\bibitem{Ghosh:2007wy} 
  S.~K.~Ghosh, T.~K.~Mukherjee, M.~G.~Mustafa and R.~Ray,
  ``PNJL model with a Van der Monde term,''
  Phys.\ Rev.\ D {\bf 77}, 094024 (2008)
  [arXiv:0710.2790 [hep-ph]]. 
  
\bibitem{Hansen:2006ee} 
  H.~Hansen, W.~M.~Alberico, A.~Beraudo, A.~Molinari, M.~Nardi and C.~Ratti,
  Phys.\ Rev.\ D {\bf 75}, 065004 (2007)
  doi:10.1103/PhysRevD.75.065004
  [hep-ph/0609116].  
  
\bibitem{Islam:2014tea} 
  C.~A.~Islam, R.~Abir, M.~G.~Mustafa, R.~Ray and S.~K.~Ghosh,
  ``The consequences of $SU(3)$ colorsingletness, Polyakov Loop and $Z(3)$ symmetry on a quark–gluon gas,''
  J.\ Phys.\ G {\bf 41}, 025001 (2014)
  [arXiv:1208.3146 [hep-ph]]. 
  
\bibitem{Ferrari:2012yw} 
  G.~N.~Ferrari, A.~F.~Garcia and M.~B.~Pinto,
  Phys.\ Rev.\ D {\bf 86}, 096005 (2012)
  doi:10.1103/PhysRevD.86.096005
  [arXiv:1207.3714 [hep-ph]].  
  
 \bibitem{Gusynin:1995nb} 
   V.~P.~Gusynin, V.~A.~Miransky and I.~A.~Shovkovy,
   Nucl.\ Phys.\ B {\bf 462}, 249 (1996)
   doi:10.1016/0550-3213(96)00021-1
   [hep-ph/9509320].  
   
 \bibitem{Schramm:1991ex} 
   S.~Schramm, B.~Muller and A.~J.~Schramm,
   Mod.\ Phys.\ Lett.\ A {\bf 7}, 973 (1992).
   doi:10.1142/S0217732392000860.
  
\bibitem{Fraga:2008qn} 
  E.~S.~Fraga and A.~J.~Mizher,
  Phys.\ Rev.\ D {\bf 78}, 025016 (2008)
  doi:10.1103/PhysRevD.78.025016
  [arXiv:0804.1452 [hep-ph]].  
  
\bibitem{Gatto:2010pt} 
  R.~Gatto and M.~Ruggieri,
  Phys.\ Rev.\ D {\bf 83}, 034016 (2011)
  doi:10.1103/PhysRevD.83.034016
  [arXiv:1012.1291 [hep-ph]].  
  
\bibitem{Bali:2012zg} 
  G.~S.~Bali, F.~Bruckmann, G.~Endrodi, Z.~Fodor, S.~D.~Katz and A.~Schafer,
  Phys.\ Rev.\ D {\bf 86}, 071502 (2012)
  doi:10.1103/PhysRevD.86.071502
  [arXiv:1206.4205 [hep-lat]].  
  
\bibitem{Bruckmann:2013oba} 
  F.~Bruckmann, G.~Endrodi and T.~G.~Kovacs,
  JHEP {\bf 1304}, 112 (2013)
  doi:10.1007/JHEP04(2013)112
  [arXiv:1303.3972 [hep-lat]]. 
  
\bibitem{Endrodi:2015oba} 
  G.~Endrodi,
  JHEP {\bf 1507}, 173 (2015)
  doi:10.1007/JHEP07(2015)173
  [arXiv:1504.08280 [hep-lat]].  
 
\bibitem{Chao:2013qpa} 
  J.~Chao, P.~Chu and M.~Huang,
  Phys.\ Rev.\ D {\bf 88}, 054009 (2013)
  doi:10.1103/PhysRevD.88.054009
  [arXiv:1305.1100 [hep-ph]].  
  
\bibitem{Andersen:2014oaa} 
  J.~O.~Andersen, W.~R.~Naylor and A.~Tranberg,
  JHEP {\bf 1502}, 042 (2015)
  doi:10.1007/JHEP02(2015)042
  [arXiv:1410.5247 [hep-ph]]. 
  
\bibitem{Pisarski:1987wc} 
  R.~D.~Pisarski,
  Nucl.\ Phys.\ B {\bf 309}, 476 (1988).
  doi:10.1016/0550-3213(88)90454-3  
  
\bibitem{Bzdak:2012fr} 
  A.~Bzdak and V.~Skokov,
  Phys.\ Rev.\ Lett.\  {\bf 110}, no. 19, 192301 (2013)
  doi:10.1103/PhysRevLett.110.192301
  [arXiv:1208.5502 [hep-ph]].  
  
\bibitem{McLerran:2013hla} 
  L.~McLerran and V.~Skokov,
  Nucl.\ Phys.\ A {\bf 929}, 184 (2014)
  doi:10.1016/j.nuclphysa.2014.05.008
  [arXiv:1305.0774 [hep-ph]].  
  
\bibitem{Roessner:2006xn} 
  S.~Roessner, C.~Ratti and W.~Weise,
  Phys.\ Rev.\ D {\bf 75}, 034007 (2007)
  doi:10.1103/PhysRevD.75.034007
  [hep-ph/0609281].  
  
\bibitem{Ferrer:2013noa} 
  E.~J.~Ferrer, V.~de la Incera, I.~Portillo and M.~Quiroz,
  Phys.\ Rev.\ D {\bf 89}, no. 8, 085034 (2014)
  doi:10.1103/PhysRevD.89.085034
  [arXiv:1311.3400 [nucl-th]].
  

\bibitem{Frolov:2010wn} 
  I.~E.~Frolov, V.~C.~Zhukovsky and K.~G.~Klimenko,
  Phys.\ Rev.\ D {\bf 82}, 076002 (2010)
  doi:10.1103/PhysRevD.82.076002
  [arXiv:1007.2984 [hep-ph]].
  

\bibitem{Ferrer:2019zfp} 
  E.~J.~Ferrer and V.~de la Incera,
  arXiv:1902.06810 [nucl-th].
   
\end{thebibliography}
\end{document}